\documentclass[%
 reprint,
superscriptaddress,
 amsmath,amssymb,
 aps,
 prl,
]{revtex4-2}
\usepackage{graphicx}
\usepackage{dcolumn}
\usepackage{bm}
\usepackage{epsfig}
\usepackage{amsmath}
\usepackage{verbatim}
\usepackage{amssymb}
\usepackage{array}
\usepackage{booktabs}
\usepackage{subcaption}
\usepackage[many]{tcolorbox}

\begin{document}

\title{Continuous Wave Quantum Detection and Ranging with quantum heterodyne detection}

\author{Ming-Da Huang}
\author{Zhan-Feng Jiang}
\author{M. Hunza}
\author{Long-Yang Cao}
\author{Hong-Yi Chen}

	\affiliation {College of Physics and Optical Engineering, Shenzhen University, Shenzhen 
        518060, China}
        \affiliation{State Key Laboratory of Radio Frequency Heterogeneous Integration, Shenzhen University, Shenzhen 518060, China}
        \affiliation{Institute of Intelligent Optical Measurement and Detection, Shenzhen University, Shenzhen 518060, China}

\author{Yuan-Feng Wang}
  \affiliation{Quantum Science Center of Guangdong-HongKong-Macao Greater Bay Area (Guangdong), Shenzhen 518045, China}

\author{Yuan-Yuan Zhao}
\affiliation{Quantum Science Center of Guangdong-HongKong-Macao Greater Bay Area (Guangdong), Shenzhen 518045, China}

\author{Hai-Dong Yuan}
\address{Department of Mechanical and Automation Engineering, The Chinese University of Hong Kong, Shatin, Hong Kong SAR, China}

\author{Qi Qin}
\email{qi.qin@szu.edu.cn}
	      \affiliation{Department of System Engineering, City University of Hong Kong, 83 Tat Chee Avenue, Kowloon, Hong Kong SAR, China}
        \affiliation {College of Physics and Optical Engineering, Shenzhen University, Shenzhen 
         518060, China}
        \affiliation{State Key Laboratory of Radio Frequency Heterogeneous Integration, Shenzhen University, Shenzhen 518060, China}
        \affiliation{Institute of Intelligent Optical Measurement and Detection, Shenzhen University, Shenzhen 518060, China}
        \affiliation{Quantum Science Center of Guangdong-HongKong-Macao Greater Bay Area (Guangdong), Shenzhen 518045, China}

\begin{abstract}

In the continuous-wave Detection and Ranging technology, simultaneous and accurate range and velocity measurements of an unknown target are typically achieved using a frequency-modulated continuous wave (FMCW) with a heterodyne receiver. The high time-bandwidth product of the FMCW waveform facilitates the optimization and high-precision of these measurements while maintaining low transmission power. Despite recent efforts to develop the quantum counterpart of this technology, a quantum protocol for FMCW that enhances measurement precision in lossy channels with background noise has yet to be established. Here, we propose a quantum illumination protocol for FMCW technology that utilizes sum frequency generation and an entangled light source with low transmission power. This protocol demonstrates a 3 dB enhancement in the precision limit for high-loss channels compared to classical approaches, independent of the background noise level. This precision limit is achieved through quantum heterodyne detection (QHD), followed by signal processing. Moreover, in classical approaches, QHD is only optimal in high-loss channels when strong background noise is present. In weak background noise scenarios, our protocol can further provides precision enhancements up to 6 dB over classical methods with QHD.

\end{abstract}
\maketitle

\begin{figure}[h]
\centering
\includegraphics[width=1\linewidth]{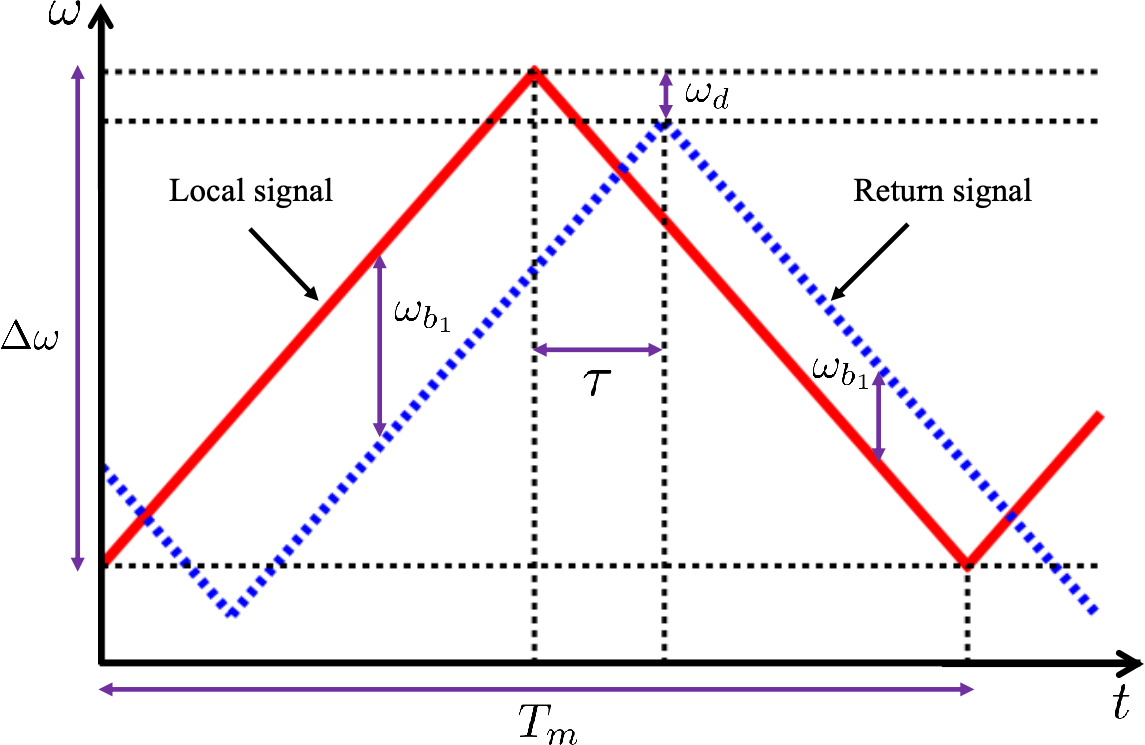}
\caption{The angular frequency $\omega(t)$ of the Local signal and the return signal under the triangle frequency modulation with initial angular frequency $\omega_{0}$, modulation bandwidth $\Delta\omega$ and modulation period $T_{m}$. Here, $\tau = \frac{T_{m}}{2\Delta \omega} \frac{\omega_{b_{1}} + |\omega_{b_{2}}|}{2}$ and $\omega_{d} = \frac{\omega_{b_{1}} - |\omega_{b_{2}}|}{2}$, where $\omega_{b_{1}}$ and $\omega_{b_{2}}$ represent the frequency differences between the received and local FMCW signals for the rising and falling edges of the triangular frequency modulation, respectively.}
\label{fig:2}
\end{figure}

\emph{Introduction.}---In Continuous-Wave (CW) Detection and Ranging systems, simultaneous measurement of a target's range and velocity is typically achieved using a Frequency-Modulated Continuous Wave (FMCW) signal combined with a heterodyne receiver that is applicable to both optical \cite{kimNanophotonicsLightDetection2021,li2022progress,li2020lidar} and microwave \cite{patole2017automotive,hakobyan2019high,deng2020silicon} regimes. The FMCW waveform is characterized by a linear frequency modulation against time, with a modulation bandwidth of $\Delta\omega$ and a modulation period of $T_{m}$, satisfying $\Delta\omega T_{m} \gg 1$. During transmission, the FMCW signal frequency varies predictably, either by increasing or decreasing against time. By comparing the frequency of the received FMCW signal with that of a local FMCW signal, both the time delay $\tau$ (which corresponds to the target's range) and the Doppler shift $\omega_{d}$ (which provides the target's velocity) can be accurately determined, as illustrated in Fig. \ref{fig:2}. This comparison is performed at the heterodyne receiver, which coherently mixes the received and local signals to produce an intermediate frequency signal. Leveraging the high time-bandwidth product of the FMCW waveform, the simultaneous measurements of $\tau$ and $\omega_{d}$ can achieve optimized resolutions that are inversely proportional to $\Delta\omega$ and $T_{m}$, respectively. The precision limit for estimating these parameters is governed by the Cramér-Rao bound (CRB), which improves as the signal-to-noise ratio (SNR) increases. For a single modulation period, it follows the relation $\text{SNR} \propto P_T T_{m}$, where $P_T$ represents the mean power of the FMCW signal. This relationship suggests that the high time-bandwidth product also enables the system to operate in a low-power regime while maintaining a high SNR, thereby ensuring high-precision detection. All of these exceptional characteristics make the system highly suitable for on-chip integration (due to low transmission power), and enable a high-resolution (since $\Delta\omega T_{m} \gg 1$), high-precision (attributable to the high SNR) detection platform, particularly advantageous for optical applications \cite{behroozpourLidarSystemArchitectures2017,riemensbergerMassivelyParallelCoherent2020,zhangLargescaleMicroelectromechanicalsystemsbasedSilicon2022,lihachevLownoiseFrequencyagilePhotonic2022}.

In recent years, the advancement of quantum information science has brought substantial focus to the concept of quantum Detection and Ranging \cite{giovannettiQuantumenhancedPositioningClock2001,giovannettiPositioningClockSynchronization2002,macconeQuantumRadar2020,lloydEnhancedSensitivityPhotodetection2008,changQuantumenhancedNoiseRadar2019,barzanjehMicrowaveQuantumIllumination2015,zhuangQuantumRangingGaussian2021,liuEnhancingLIDARPerformance2019,reichertQuantumenhancedDopplerLidar2022,shapiroQuantumPulseCompression2007,zhuangUltimateAccuracyLimit2022,zhuangEntanglementenhancedLidarsSimultaneous2017,huangQuantumLimitedEstimationRange2021,blakey2022quantum,mrozowski2024demonstration}. Notably, quantum counterparts of FMCW technologies have been proposed \cite{huang2025frequency}, offering enhanced precision and resolution. However, these enhancements are generally restricted to low-loss channels, while in high-loss channels, quantum detection techniques yield improvements primarily in resolution \cite{huang2024quantum}. Here, our focus is on applying quantum FMCW technologies to high-loss channels with varying levels of background noise in high-SNR scenarios. This approach is well-suited for real-world implementations in both the microwave regime, characterized by significant background noise ($\sim$100s to 1000s of photons per mode), and the optical regime, where the background noise is negligible ($\ll 1$ photon per mode). In both cases, the power returned from an unresolved target at a range $d$ decreases inversely with $d^{4}$ \cite{richards2005fundamentals}, resulting in substantial propagation loss. 

While quantum illumination integrated with FMCW techniques has yet to be explored, recent advancements for quantum illumination ranging in the microwave regime without FMCW \cite{zhuangUltimateAccuracyLimit2022} has shown a 3 dB improvement in precision limits over classical approaches in high-SNR scenarios for high-loss channels with significant background noise. This underscores the potential of quantum illumination to improve resilience against loss and noise in quantum Detection and Ranging protocols.

In this letter, we  propose a quantum illumination protocol for FMCW technology in high-SNR scenarios. This protocol harnesses the entanglement of in-phase and quadrature (I/Q) components along with the classical frequency correlation between the FMCW signal mode and the retained FMCW idler mode. Remarkably, our analysis shows that in high-loss channels with weak background noise, the quantum CRB of two-mode squeezed state with frequency modulation and low transmission power is lowered by introducing sum frequency generation (SFG), while it remains unchanged in high-loss channels with strong background noise. As a result, our protocol achieves a 3 dB enhancement in the precision limit for high-loss channels compared to a classical FMCW Detection and Ranging system with identical modulation bandwidth, modulation period, and low transmitted energy, regardless of background noise level. Furthermore, we find that the quantum heterodyne detection (QHD) is an optimal strategy for sufficiently strong SFG strength, enabling the classical CRB to match the quantum CRB. The classical CRB can be achieved through forward data processing, specifically using a discrete Fourier transform (DFT). Interestingly, we discover that, in classical systems, QHD is the optimal strategy in high-loss channels when strong background noise is present. Consequently, our protocol provides precision enhancements of 6 dB and 3 dB over classical methods in weak and strong background noise scenarios, respectively.

\emph{Quantum description of ranging and velocity measurement for FMCW.}---
Consider a CW field with frequency $\omega$, which is represented by a quantum state $\hat{\rho}_{S}\left(\hat{a}_{S}(\omega),\hat{a}_{S}^{\dag}(\omega)\right)$ with conjugate field operators, where the field occupies a single spatial and frequency mode. The frequency modulation process for this field can be expressed as
\begin{equation}
\hat{a}_{S}(\omega)\rightarrow\frac{1}{\sqrt{j_B+1}} \sum_{p=p_0}^{p_0+j_B}  E_{\Delta \omega_{S}}(t_p,\omega_{S})\hat{a}_{S}\left(t_p\right),
 \end{equation}
where, $E_{\Delta \omega_{S}}(t_{p},\omega_{S})$ represents the FMCW waveform with a modulation bandwidth $\Delta\omega_{S}$ and a central frequency $\omega_{S}$. The normalized field operator for this multi-frequency-mode CW field in time domain is given by
\begin{equation}
\hat{a}_{S}(t_{p}) =\frac{1}{\sqrt{j_{B}+1}} \sum_{j=j_{c}-\frac{j_{B}}{2}}^{j_{c}+\frac{j_{B}}{2}} \hat{a}_{S}\left(\omega_{j}\right) \exp \left(-i \omega_{j} t_{p}\right),
\end{equation}
with $\omega_{j}=2\pi j/T_{m}$, satisfying $\hat{a}_{S}(t_{p})=\hat{a}_{S}(t_{p}+T_{m})$ for a modulation period $T_{m}$. 
Here, the bandwidth of the field is limited to $\Delta\omega_{B}=2\pi j_{B}/T_{m}\geq \Delta\omega_{S}$, and a period of modulation $T_{m}$ is divided into $j_{B}$ intervals, with 
$t_{p}=pT_{m}/j_{B}$ where $p\in\left\{p_{0},p_{0}+1,p_{0}+2, \ldots p_{0}+j_{B}\right\}$ and $p_{0}$ is the initial time point. For such signal, the mean photon number in a single modulation period is given by $N_{S}=\sum^{p_{0}+j_{B}}_{p=p_{0}} \langle \hat{a}^{\dag}_{S}(t_{p}) \hat{a}_{S}(t_{p}) \rangle$. Additional details on quantum light fields with frequency modulation are mentioned in Sec. I of the Supplementary Material.

If this signal is sent to an unresolved target located at a range $d$ having velocity $v$, the FMCW waveform of the return field will transform into $E_{\Delta \omega_S}\left(t_p-\tau, \omega_S-\omega_{d}\right)$, along with the return field operator given as
\begin{equation}
\hat{a}_R(t_{p})=\sqrt{\epsilon} \hat{a}_S(t_{p})+\sqrt{1-\epsilon} \hat{a}_{th}(t_{p})
\end{equation}
where $\tau=2d/c$ is the round-trip time delay and $\omega_d=2\omega_{S}v/c$ is the Doppler shift. Here, the target is modeled as a fictitious beam splitter with reflectivity $\epsilon$, which can also be interpreted as the transmissivity of the propagation channel, where $\epsilon \ll 1$ for high-loss channels. The background noise with normalized field operator
\begin{equation}
\hat{a}_{th}(t_{p}) =\frac{1}{\sqrt{j_{B}+1}} \sum_{j=j_{c}-\frac{j_{B}}{2}}^{j_{c}+\frac{j_{B}}{2}} \hat{a}_{th}\left(\omega_{j}\right) \exp \left(-i \omega_{j} t_{p}\right),
\end{equation}
is modeled as a multi-frequency-mode CW thermal state with a bandwidth $\Delta\omega_{B}$ and a mean photon number $n_j=1 /\left[\exp \left(\hbar \omega_j / k_B T_{th}\right)-1\right]$ for each mode where $\hbar$ is the reduced Planck constant, $k_B$ the Boltzmann constant, and $T_{th}$ is the noise temperature. Furthermore, by comparing the waveform of the return field with that of the local field, $E_{\Delta \omega_L}\left(t_p, \omega_L\right)$, through QHD, the information regarding the range and velocity of the target can be extracted simultaneously, as described in Fig. (\ref{fig:2}).

\emph{QHD followed by SFG.}---Due to the frequency difference between the local and return fields in QHD, the image-band field $\hat{a}_{IB}$, which is the spectral counterpart of the return field mirrored around the local field frequency, inevitably combines with the return field, acting as a vacuum fluctuation \cite{collett1987quantum}. This additional vacuum fluctuation allows for the simultaneous measurement of the I/Q components \cite{shapiro2009quantum,kikuchi2015fundamentals} of the return field, albeit with a lower SNR, as discussed in Sec. IIA of the supplementary material. For a multi-frequency-mode CW field with a strong local field, the I/Q components of the return field in the time domain, as measured by QHD, are given by
\begin{equation}
\begin{aligned}
\hat{I}_{R}(t_{p})&= \frac{1}{2}[\hat{a}_{R}(t_{p})e^{i \phi_{L_{R}}(t_{p})}+\hat{a}_{R}^{\dag}(t_{p})e^{-i \phi_{L_{R}}(t_{p})}],\\
\hat{Q}_{R}(t_{p})&= -\frac{i}{2}[\hat{a}_{R}(t_{p})e^{i \phi_{L_{R}}(t_{p})}-\hat{a}_{R}^{\dag}(t_{p})e^{-i \phi_{L_{R}}(t_{p})}],
\end{aligned}
\end{equation}
respectively, satisfying $[\hat{I}_{R}(t_{p}),\hat{Q}_{R}(t_{p})]=\frac{i}{2}$, where $e^{i \phi_{L_{R}}(t_{p})}$ represents the waveform of the local field for the return field, which can take any form, including FMCW or single-frequency waveforms.
 
Furthermore, SFG can be applied to the return and idler fields prior to the QHD to measure the cross-correlation of their I/Q components \cite{zhuang2017optimum}. An SFG process can be viewed as the time-reversed counterpart of spontaneous parametric down-conversion (SPDC) \cite{o2009time,liu2020joint}. In the limit of an infinitely wide SPDC photon bandwidth, the SFG process, denoted by $\hat{U}_{S F G}$, applied to the return and idler fields in the time domain, can be approximated as
\begin{equation}\label{equ9}
\begin{aligned}
\hat{I}_{s}+\epsilon_s \sqrt{j_B+1}\sum_{p=p_{0}}^{p_0+j_B}\left[\hat{a}_P^{\dag}\left(t_p\right) \hat{a}_R\left(t_p\right) \hat{a}_I\left(t_p\right)-H.C.\right],
\end{aligned}
\end{equation}
under the condition, when the mean photon number per time unit in the idler mode is sufficiently small, where $\hat{I}_{s}$ is the identity operator and $\epsilon_s$ represents the strength of the SFG process. 
Here, $P$ refers to the up-conversion mode (or pump mode in the case of SPDC), and $I$ denotes the idler mode. Under these conditions, there is a clear relationship between the I/Q components of the field before and after the SFG process given as
\begin{equation}\label{equ2}
\begin{aligned}
\hat{U}_{SFG}^{\dag} &\hat{I}_P(t_{p}) \hat{U}_{SFG} \propto \epsilon_s\left[\hat{I}_R\left(t_p\right) \hat{I}_I\left(t_p\right)-\hat{Q}_R\left(t_p\right) \hat{Q}_I\left(t_p\right)\right],\\
\hat{U}_{SFG}^{\dag} &\hat{Q}_P(t_{p}) \hat{U}_{SFG}\propto \epsilon_s \left[\hat{I}_R\left(t_p\right) \hat{Q}_I\left(t_p\right)+\hat{Q}_R\left(t_p\right) \hat{I}_I\left(t_p\right)\right],
\end{aligned}
\end{equation}
if $\phi_{L_{P}}(t_p)=\phi_{L_{R}}(t_p)+\phi_{L_{I}}(t_p)$ and $\epsilon_s\neq 0$, as discussed in Sec. IIB of the Supplementary Material. This relationship suggests that, under the approximation of the up-conversion unitary operator $\hat{U}_{SFG}$ as shown in Eq. (\ref{equ9}), measuring the I/Q components of the up-converted field after the SFG process is equivalent to measuring the sum and difference in covariances between the I/Q components of the return-idler field, $\hat{\rho}_{R,I}$, prior to the SFG process.

\begin{figure*}[!ht]
    \centering
    
        \begin{tabular}{|c|c|c|} 
            \hline
            \begin{subfigure}[t]{0.35\textwidth}
                \centering
                \includegraphics[width=\textwidth]{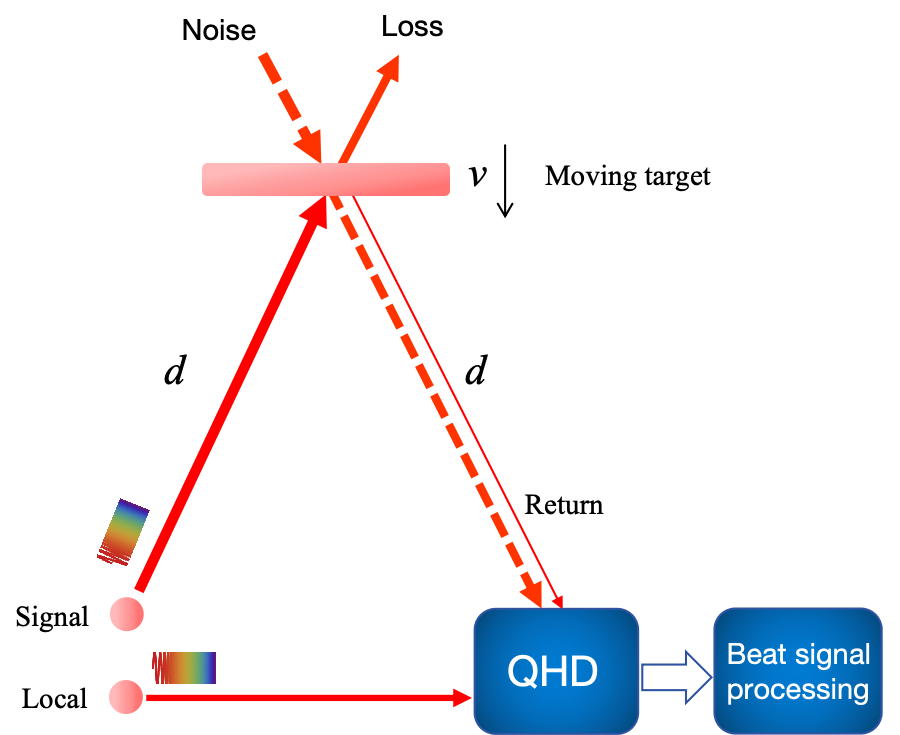}
                \caption{FMCW classical illumination}
                \label{fig:CS_FMCW}
            \end{subfigure} &
            \begin{subfigure}[t]{0.28\textwidth}
                \centering
                \includegraphics[width=\textwidth]{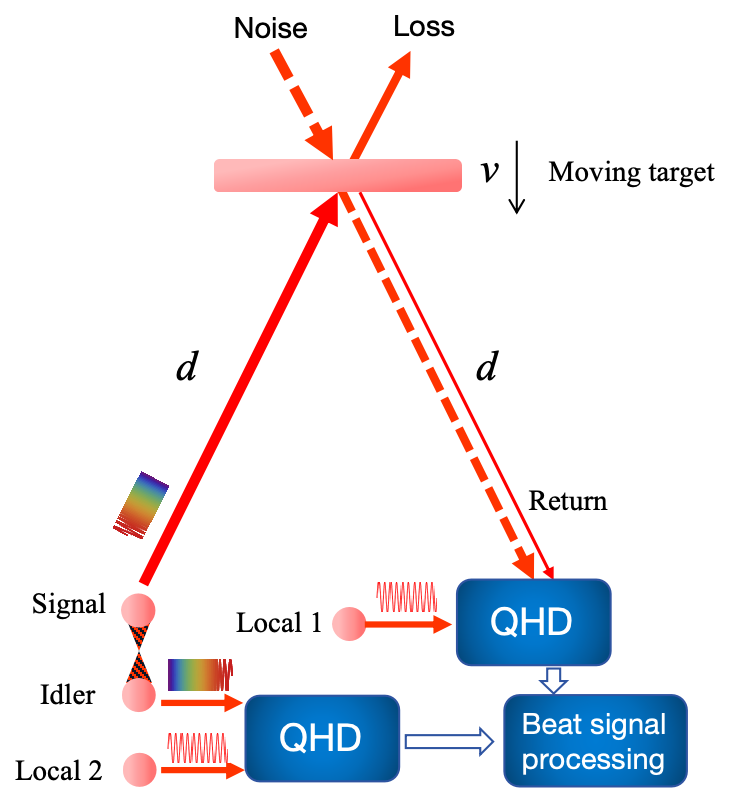}
                \caption{FMCW quantum illumination with independent QHD}
                \label{fig:TMSS_FMCW}
            \end{subfigure} &
            \begin{subfigure}[t]{0.35\textwidth}
                \centering
                \includegraphics[width=\textwidth]{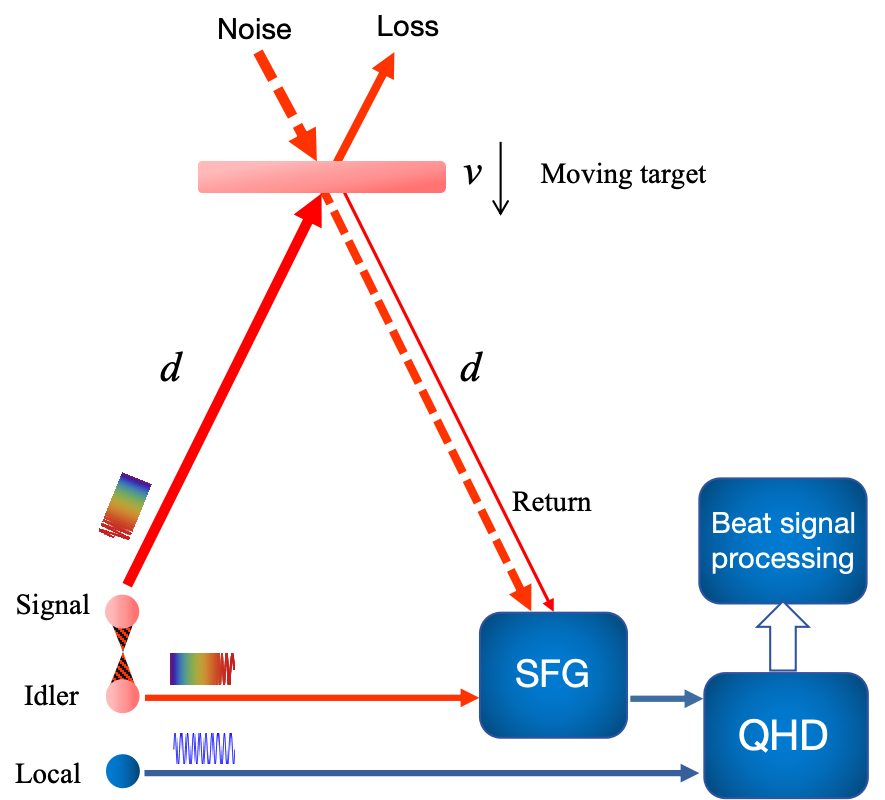}
                \caption{FMCW quantum illumination with SFG}
                \label{fig:TMSS_SFG_FMCW}
            \end{subfigure} \\
            \hline
        \end{tabular}
    
    \caption{Sketches of various FMCW illumination for simultaneous range and velocity measurement.}
    \label{fig:FMCW illumination}
\end{figure*}

\emph{FMCW classical illumination.}---As a benchmark, Fig. \ref{fig:FMCW illumination} (a) illustrates a classical illumination scheme utilizing a FMCW laser beam. The return field of this scheme is a mixed state, consisting of a FMCW coherent state and a CW thermal state, $\hat{\rho}_{R}=\hat{\rho}_{coh-th}$, and is fully described by its Wigner function in the form of a Gaussian distribution. This distribution encapsulates the mean values and covariance of the field's I/Q components in the time domain, expressed as
\begin{equation}\label{equ8}
\begin{aligned}
\left\langle \hat{I}_R(t_{p})\right\rangle_{coh-th}=\sqrt{\epsilon n_{coh}} \cos \left(\omega_l t_{p}+\phi_l\right),\\
\left\langle \hat{Q}_R(t_{p})\right\rangle_{coh-th}=\sqrt{\epsilon n_{coh}}\sin \left(\omega_l t_{p}+\phi_l\right),
\end{aligned}
\end{equation}
and 
\begin{equation}
C_{coh-th}=\epsilon C_{coh}+(1-\epsilon) C_{th},
\end{equation}
where $C_{coh}=I_{2}/4$ and $C_{th }=(1+2 n_{th})I_{2}/4$ with two-dimensional identity matrix $I_{2}$. Here, triangular frequency modulation waveform, as depicted in Fig. \ref{fig:2}, is applied to both the signal and local field, resulting in a beat frequency $\omega_l$ and phase $\phi_l$, where $l=1$ and $l=0$ correspond to the first and second halves of the modulation period, respectively. The exact form of $\omega_l$ and $\phi_l$ can be found in the Sec. IIIA of the supplementary material. Moreover, the mean photon number per temporal mode is given by $n_{coh}=|\alpha|^2/(j_{B}+1)$ for the FMCW coherent state, and $n_{th}\equiv \sum_{j} n_{j}/(j_{B}+1)$ for the CW thermal state.
Thus, the instantaneous quantum Fisher information (QFI) of the beat frequency $\omega_l$ for this Gaussian state \cite{weedbrook2012gaussian,nichols2018multiparameter}, which represents the inverse of its quantum CRB, is given by
\begin{equation}\label{equ1}
F_{coh-th}^{Q}(\omega_l,t_{p})\approx\frac{4 t_{p}^2 n_{coh} \epsilon}{1+2 n_{t h}},
\end{equation}
where the approximation holds for a high-loss channel with $\epsilon\ll 1$, as shown in Sec. IIIC of the supplementary material.

Further, the instantaneous classical Fisher information (CFI) for QHD, $F^{QHD}_{c o h-t h}\left(\omega_l, t_{p}\right)$, which represent the inverse of classical CRB, can be evaluated in comparison with the QFI. As shown in Sec. IVA of the supplementary material, under a high-loss channel ($\epsilon \ll 1$) with weak background noise ($n_{th} \ll 1$), it is found that $F^{QHD}_{coh-th}\left(\omega_l, t_{p}\right) = F^Q_{coh-th}\left(\omega_l, t_{p}\right)/2$, due to the vacuum noise of image band.  Conversely, for a high-loss channel ($\epsilon\ll 1$) with strong background noise ($n_{th}\gg1$), $F^{QHD}_{c o h-t h}\left(\omega_l, t\right)=F^Q_{c o h-t h}\left(\omega_l, t_{p}\right)$. These results indicate that QHD is an optimal detection strategy for FMCW classical illumination in the presence of strong background noise.

\emph{FMCW quantum illumination with independent QHD.}---A narrow band two-path-mode squeezed vacuum state with frequency modulation, which can be generated by Cavity-Enhanced SPDC process \cite{slattery2019background} with frequency modulation, is used to construct a FMCW quantum illumination scheme. One path acts as the signal mode, with waveform $E_{\Delta \omega_S}\left(t_p, \omega_S\right)$ and mean photon number $n_{sv}$ per temporal mode, sent toward the target and mixed with background noise. The other serves as the idler mode, with waveform $E_{\Delta \omega_I}\left(t_p, \omega_I\right)$ and the same mean photon number $n_{sv}$, retained locally. Moreover, the central frequencies of the signal and idler mode satisfy the relation $\omega_{P}=\omega_{S}+\omega_{I}$, as they are produced by the SPDC process, where $\omega_{P}$ is the frequency of the pump mode. In this case, the return-idler field, $\hat{\rho}_{R,I}=\hat{\rho}_{sv-th}$, remains a Gaussian state with mean values $\left\langle I_R(t)\right\rangle_{sv-th}=\left\langle Q_R(t)\right\rangle_{sv-th}= \left\langle I_I(t)\right\rangle_{sv-th}=\left\langle Q_I(t)\right\rangle_{sv-th}=0$, and a covariance matrix given by 
\begin{equation}\label{equ3}
C_{sv-th}=\left(\begin{array}{cc}
\epsilon C_{sv}+ (1-\epsilon)C_{th} &\sqrt{\epsilon}\Lambda^{\prime}(\omega_{l},\phi_l) \\
\sqrt{\epsilon} \Lambda^{\prime}(\omega_{l},\phi_l) &C_{sv}
\end{array}\right),
\end{equation}
where $C_{sv}=(1+2 n_{s v})I_{2}/4$, and $\Lambda^{\prime}(\omega_{l},\phi_l)=-\frac{\sqrt{n_{s v}\left(1+n_{s v}\right)}}{2}[\cos \left(\omega_l t_p+\phi_l\right) \sigma_{z} +\sin \left(\omega_l t_p+\phi_l\right) \sigma_{x}  ]$ with $\sigma_{z}$ and $\sigma_{x}$ being Pauli matrix. Here, we introduce a classical correlation by setting $\Delta \omega_S=-\Delta \omega_I$ ensuring that the sum frequency of the signal and idler modes remains constant. This suggests that if the frequency of the signal mode increases linearly with time, the frequency of the idler mode decreases linearly with time, and vice versa.
Furthermore, the local fields for both the return and idler modes are chosen to be single-frequency waveforms, satisfying $\omega_{L_{R}}+\omega_{L_{I}}=\omega_S+\omega_I=\omega_{P}$. The matrix $C_{sv-th}$ clearly demonstrates that the covariance oscillates at the frequency $\omega_{l}$. Thus, the instantaneous QFI of the beat frequency $\omega_l$ for this Gaussian state can be given as
\begin{equation}\label{equ4}
F^{Q}_{sv-th}(\omega_l,t_{p})\approx\frac{4 t_{p}^2 n_{sv}\epsilon}{1+n_{th}},
\end{equation}
where the approximation holds for a high loss channel $\epsilon\ll 1$ with low transmission power $n_{sv}\ll1$, as shown in Sec. IIIC of the supplementary material. By comparing it with Eq. (\ref{equ1}) under the same transmission power $n_{sv}=n_{coh}$, we find that they are identical in the regime of weak background noise $n_{th}\ll1$.  However, in the presence of strong background noise $n_{th}\gg1$, $F_{sv-th}^{Q}(\omega_l,t_{p})=2F_{coh-th}^{Q}(\omega_l,t_{p})$, indicating a 3 dB improvement in the precision limit for FMCW quantum illumination.

As illustrated in Fig. \ref{fig:FMCW illumination} (b), a straightforward detection strategy for this quantum FMCW illumination is to apply QHD independently to both the signal and idler modes of the returned state. This detection strategy measures the covariance matrix $C_{sv-th}$, similar to the proposal presented in Ref. \cite{changQuantumenhancedNoiseRadar2019}. Interestingly, under the same transmission power $n_{sv}=n_{coh}$, we find that the instantaneous CFIs satisfy $F^{QHD}_{sv-t h}\left(\omega_l, t_p\right)=F^{QHD}_{c o h-t h}\left(\omega_l, t_p\right)$, as shown in Sec. IVA of the supplementary material. This indicates that, despite the quantum CRB improvement, the classical CRB remains identical for both quantum and classical illuminations, as illustrated in Fig. \ref{fig:FMCW illumination} (a) and (b).

\emph{FMCW quantum illumination with SFG.}---To further demonstrate quantum enhancement, a SFG process can be applied to the signal and idler modes of $\rho_{sv-th}$ prior to the measurement, as shown in Fig. \ref{fig:FMCW illumination} (c). Since SFG is the reverse process of SPDC, it can coherently preserve the information of the return-idler field, which has been shown to be useful in quantum illumination for both theoretical \cite{zhuang2017optimum,liu2020joint} and experimental \cite{liu2023compact} studies.

In this scenario, according to Eq. (\ref{equ2}), the SFG process combines the cross-correlation term in the covariance matrix of the return-idler field, as shown in Eq. (\ref{equ3}), when the local field is chosen to be a single-frequency waveform satisfying $\omega_{L} = \omega_S + \omega_I = \omega_{P}$, and the classical correlation $\Delta \omega_S = -\Delta \omega_I$ is applied. Here, the SFG process acts as a quantum matched filter that decodes the information embedded in the signal field, yielding an up-converted state $\rho_{SFG}$ corresponding to a single-frequency beam with a mean value as shown below
\begin{equation}\label{equ5}
\begin{aligned}
\left\langle I_P(t_{p})\right\rangle_{SFG}=-\epsilon_s\sqrt{\epsilon n_{s v}} \cos \left(\omega_l t_{p}+\phi_l\right),\\ 
\left\langle Q_P(t_{p})\right\rangle_{SFG}=-\epsilon_s\sqrt{\epsilon n_{s v}} \sin \left(\omega_l t_{p}+\phi_l\right),
\end{aligned}
\end{equation}
and a covariance matrix 
\begin{equation}
C_{SFG}\approx\frac{1}{8}\epsilon^{2}_s\left(\begin{array}{cc}
1+2 n_{t h} & 0 \\
0 & 1+2 n_{t h}
\end{array}\right),
\end{equation}
where the approximation holds for $\epsilon \ll 1$ and $n_{s v} \ll 1$, that is derived from Eq. (\ref{equ2}) and Eq. (\ref{equ3}). In this case, the instantaneous QFI of the beat frequency $\omega_l$ can be approximated as
\begin{equation}\label{equ10}
F^{Q}_{SFG}(\omega_l,t_{p})\approx\frac{8 t_{p}^2n_{sv}  \epsilon}{1+2 n_{t h}},
\end{equation}
as shown in Sec. IIIC of the supplementary material. By comparing it with Eq. (\ref{equ1}) and Eq. (\ref{equ4}) under the condition of equal transmission power $n_{sv}=n_{coh}$, we find that $F^Q_{S F G}\left(\omega_l, t_{p}\right)=F^Q_{s v-t h}\left(\omega_l, t_p\right)=2F^Q_{c o h-t h}\left(\omega_l, t_p\right)$ for strong background noise $(n_{th}\gg1)$, while $F^Q_{S F G}\left(\omega_l, t\right)=2F^Q_{s v-t h}\left(\omega_l, t_p\right)=2F^Q_{c o h-t h}\left(\omega_l, t_p\right)$ for weak background noise $(n_{th}\ll1)$. These results indicate a consistent 3 dB improvement in the precision limit for FMCW quantum illumination using SFG, regardless of the background noise level, highlighting the role of SFG as an effective quantum noise suppressor in certain situations.

Following the SFG process, QHD is performed on the up-converted state, producing a beat signal analogous to that of FMCW classical illumination. As shown in Sec. IVA of the supplementary material, for $\epsilon \ll 1$ and $n_{s v}=n_{coh} \ll 1$, the instantaneous CFI for QHD, can be approximated as
\begin{equation}\label{equ6}
F_{SFG}^{QHD}(\omega_{l},t_{p})\approx\frac{8t_{p}^2  \epsilon n_{s v} \epsilon_s^2}{2+\left(1+2 n_{t h}\right) \epsilon_s^2},
\end{equation}
which is relate to the SFG strength $\epsilon_s$ due to the vacuum noise of image band. When the SFG strength is $\epsilon_s=\sqrt{2}$, QHD becomes the optimal method for the up-converted state $\rho_{SFG}$, achieving $F_{SFG}^{QHD}(\omega_{l},t_{p})  = F_{SFG}^{Q}(\omega_{l},t_{p})$. This conclusion can be extended to the regime where $\epsilon_s\gg\sqrt{2}$. Notably, in the region where $\epsilon^{2}_s n_{th}\gg1$ and $\epsilon_s\ll\sqrt{2}$, the relation $F_{SFG}^{QHD}(\omega_{l},t_{p})=F_{sv-th}^{Q}(\omega_{l},t_{p})$ holds.

\begin{figure}[b]
\centering
\includegraphics[width=1\linewidth]{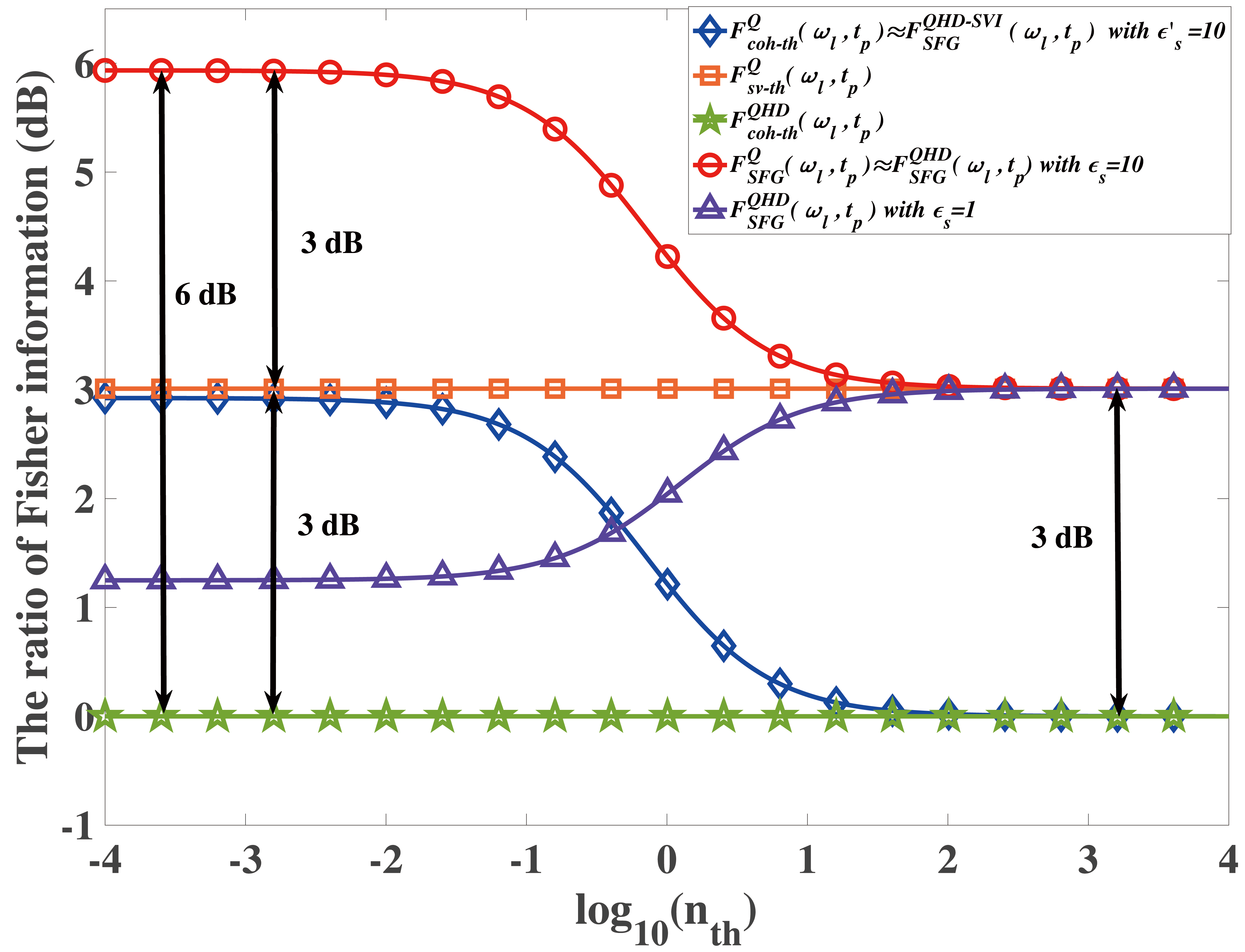}
\caption{Comparison of QFIs and CFIs for FMCW illumination at different noise levels.}
\label{fig:1}
\end{figure}

\emph{Results.}---A comparison of QFIs and CFIs for FMCW illumination with equally low transmission power at varying noise levels in a high-loss channel is shown in Fig. \ref{fig:1}, where the CFI of QHD using an FMCW coherent state serves as the benchmark. It demonstrates a 3 dB QFI improvement for FMCW quantum illumination with SFG, independent of the background noise level, as discussed after Eq. (\ref{equ10}).
Additionally, it shows that when the SFG strength is high $(\epsilon_s\gg\sqrt{2})$,
\begin{equation}
F^{QHD}_{SFG}(\omega_l,t_{p})\approx F^{Q}_{SFG}(\omega_l,t_{p})=4F^{QHD}_{coh-th}(\omega_l,t_{p}),
\end{equation}
in the regime of weak background noise ($n_{th}\ll1$) demonstrating a 6 dB  CFI improvement for FMCW quantum illumination with SFG. Conversely for strong background noise ($n_{th}\ll1$), we obtain
\begin{equation}
F^{QHD}_{SFG}(\omega_l,t_{p})\approx F^{Q}_{SFG}(\omega_l,t_{p})=2F^{QHD}_{coh-th}(\omega_l,t_{p}),
\end{equation}
corresponding to a 3 dB CFI improvement.

For low SFG strength, which is typical in experiment \cite{liu2020joint,liu2023compact}, a 3 dB improvement is observed only in the presence of strong background noise, as the line with triangle marker shown in Fig. \ref{fig:1}. This is because, for small $\epsilon_s$ and low background noise ($n_{th} \ll 1$), the signal of the up-conversion mode is modified by $\epsilon_s$, as shown in Eq. (\ref{equ5}), while the noise is dominated by the vacuum fluctuations of the image band, as shown in Eq. (\ref{equ6}), resulting in a poor SNR. As discussed in Sec. IVB of the supplementary material, one potential solution to this issue is to mitigate vacuum fluctuations by applying squeezing to the image band mode, that is, a QHD with squeezed vacuum injection (SVI). Assuming the squeezing amplitude is $r^{\prime}$ with a zero squeezing angle (i.e., squeezing the in-phase component of the image band mode), for $\epsilon \ll 1$ and $n_{s v}=n_{coh} \ll 1$, the instantaneous CFI of FMCW quantum illumination with SFG can be approximated as
\begin{equation}\label{equ7}
F_{SFG}^{QHD-SVI}(\omega_{l},t_{p})\approx\frac{4t_{p}^2  \epsilon n_{s v} \epsilon_s^{\prime 2}}{2+\left(1+2 n_{t h}\right) \epsilon_s^{\prime 2}},
\end{equation}
where the modified strength of the SFG process $\epsilon_s^{\prime}\equiv \epsilon_s e^{r^{\prime}}$ is improved by the factor of $e^{r^{\prime}}$. Here, only the CFI of the in-phase component measurement has been considered, as the noise in the quadrature component increases significantly for large squeezing amplitudes, $r^{\prime}$. For $\epsilon_s^{\prime} \gg \sqrt{2}$, $F_{SFG}^{QHD-SVI}(\omega_{l},t_{p}) = F_{coh-th}^{Q}(\omega_{l},t_{p})$, showing a 3 dB improvement over FMCW classical illumination with QHD for weak background noise $(n_{th}\ll 1)$, as the line with diamond marker shown in Fig. \ref{fig:1}.

\emph{Discussion.}---The CRB represents the precision limit in the high-SNR regime, while the QFIs and CFIs we derived apply to each temporal mode, assuming low transmission power ($n_{sv} \ll 1$) for FMCW quantum illumination. To achieve high SNR, the QFIs and CFIs over the entire modulation period should be considered. Specifically, $F(\omega_{l}) = \int_{T_{m}} dt F(\omega_{l}, t) \propto n_{sv} T_{m} = N_{S}$ in the continuous-time limit ($t_p\rightarrow t$ for sufficient large $j_{B}$). Here, the mean photon number of the signal mode over a single modulation period satisfies $N_{S} \gg n_{th}$ for strong background noise and $N_{S} \gg 1$ for weak background noise. Furthermore, the precision limits set by classical CRBs can be achieved through the DFT of the beat signal (e.g., Eq. (\ref{equ5}) and Eq. (\ref{equ8})), as the DFT serves as a maximum likelihood estimator for the beat frequency \cite{rifeSingleToneParameter1974, erkmenMaximumlikelihoodEstimationFrequencymodulated2013,richards2005fundamentals}. Subsequently, the range and velocity of the target can be simultaneously extracted from the beat frequency, just as in classical FMCW Detection and Ranging.

In conclusion, we introduce a quantum illumination protocol for FMCW technology that leverages SFG, comparing its performance to conventional FMCW classical illumination. Our protocol achieves a 3 dB improvement in the precision limit for high-loss channels, irrespective of the background noise levels. The precision limit is attained through QHD followed by DFT, yielding precision enhancements of up-to 6 dB. Even for low SFG process strength, a 3 dB precision enhancement can be observed by using QHD with SVI for weak background noise. As such, our protocol is particularly well suited for real-world implementations of simultaneous range and velocity measurement for both the microwave and optical regimes.

Finally, we emphasize that the SFG process used in FMCW quantum illumination cannot be trivially considered as part of the detection strategy, as the QFIs are altered before and after the SFG. Therefore, the SFG process not only acts as a quantum match filter to decode the information within the signal field but can also function as a quantum noise suppressor in certain situations. This provides new insights for the design of quantum illumination with SFG.

\section*{ACKNOWLEDGMENTS}

The work was supported by Guangdong Provincial Quantum Science Strategic Initiative
(GDZX2306001,GDZX2303001), Shenzhen Fundamental research
project (Grant No. JCYJ20241202123903005), and the startup fund of Shenzhen City.

\bibliography{bibtex}

\end{document}

\title{Continuous Wave Quantum Detection and Ranging with quantum heterodyne
detection: Supplementary material}

\author{Ming-Da Huang}
\author{Zhan-Feng Jiang}
\author{M. Hunza}
\author{Long-Yang Cao}
\author{Hong-Yi Chen}

	\affiliation {College of Physics and Optical Engineering, Shenzhen University, Shenzhen 
        518060, China}
        \affiliation{State Key Laboratory of Radio Frequency Heterogeneous Integration, Shenzhen University, Shenzhen 518060, China}
        \affiliation{Institute of Intelligent Optical Measurement and Detection, Shenzhen University, Shenzhen 518060, China}

\author{Yuan-Feng Wang}
  \affiliation{Quantum Science Center of Guangdong-HongKong-Macao Greater Bay Area (Guangdong), Shenzhen 518045, China}

\author{Yuan-Yuan Zhao}
\affiliation{Quantum Science Center of Guangdong-HongKong-Macao Greater Bay Area (Guangdong), Shenzhen 518045, China}

\author{Hai-Dong Yuan}
\address{Department of Mechanical and Automation Engineering, The Chinese University of Hong Kong, Shatin, Hong Kong SAR, China}

\author{Qi Qin}
\email{qi.qin@szu.edu.cn}
	      \affiliation{Department of System Engineering, City University of Hong Kong, 83 Tat Chee Avenue, Kowloon, Hong Kong SAR, China}
        \affiliation {College of Physics and Optical Engineering, Shenzhen University, Shenzhen 
         518060, China}
        \affiliation{State Key Laboratory of Radio Frequency Heterogeneous Integration, Shenzhen University, Shenzhen 518060, China}
        \affiliation{Institute of Intelligent Optical Measurement and Detection, Shenzhen University, Shenzhen 518060, China}
        \affiliation{Quantum Science Center of Guangdong-HongKong-Macao Greater Bay Area (Guangdong), Shenzhen 518045, China}

\begin{abstract}

\end{abstract}
\pacs{}
\maketitle

\section{Quantum light field with frequency modulation}

\subsection{A multi-frequency-mode continuous-wave quantum light field in discrete time domain}

For a multi-frequency-mode continuous-wave (CW) quantum light field with single spatial mode field, an annihilation operator $\hat{A}(t)$ at a certain time $t$ can be described by 
\begin{equation}
\hat{A}(t) = \sum_{j} \hat{a}(\omega_{j}) \exp{(-i \omega_{j} t)},
\end{equation}
    satisfying periodic relationship $\hat{A}(t+T_{m})=\hat{A}(t)$, where $\hat{a}(\omega_{j})$ is the annihilation operator of the light field having frequency $\omega_{j}=2\pi j/T_{m}$. 
Due to the periodicity of $\hat{A}(t)$, without loss of generality below we will restrict 
$t\in [t_{0},t_{0}+T_{m}]$. 

Since $\left[\hat{a}\left(\omega_{j}\right), \hat{a}^{\dag}\left(\omega_{j^{\prime}}\right)\right]=\delta_{j, j^{\prime}}$, we have
\begin{equation}
\left[\hat{A}(t), \hat{A}^{\dag}\left(t^{\prime}\right)\right]=\sum_{j} \exp \left[i \omega_{j}\left(t^{\prime}-t\right)\right],
\end{equation}
where $t,t’\in [t_{0},t_{0}+T_{m}]$, which do not necessarily commute when $t\neq t’$. However, under a finite bandwidth and a discrete time approximation \cite{huang2025frequency}, these commutation relations become 
\begin{equation}\label{equ1}
\left[\hat{A}(t_{p}), \hat{A}^{\dag}\left(t_{p'}\right)\right]=\left(j_{B}+1\right) \delta_{p, p'},
\end{equation}
commuting with each other when $p\neq p’$, where $j_{B}+1$ is the dimension of the discrete time as well as the discrete frequencies within the bandwidth. These approximation involves dividing the period of multi-mode CW light field, $T_m$, into $j_B$ intervals with
\begin{equation}
t_{p}=\frac{T_{m}}{j_{B}} p, \quad p\in\left\{p_{0},p_{0}+1,p_{0}+2, \ldots p_{0}+j_{B}\right\},
\end{equation}
while the bandwidth of the frequency is limited to $\Delta\omega_{B}\equiv 2\pi j_{B}/T_{m}$, i.e., the discrete frequencies are
\begin{equation}
\omega_{j}=\frac{2\pi j}{T_{m}}, \quad j\in\left\{j_{c}-\frac{j_{B}}{2}, \ldots ,j_{c}, \ldots ,j_{c}+\frac{j_{B}}{2}\right\},
\end{equation}
where $\omega_{j_{c}}=2\pi j_c/T_{m}$ represents a frequency that is closest to the central frequency of the multi-mode CW light field. 

The commutation relation mentioned in Eq. (\ref{equ1}) can be normalized by letting
\begin{equation}\label{equ2}
\hat{a}(t_{p}) \equiv \frac{\hat{A}(t_{p})}{\sqrt{j_{B}+1}}=\frac{1}{\sqrt{j_{B}+1}} \sum_{j=j_{c}-\frac{j_{B}}{2}}^{j_{c}+\frac{j_{B}}{2}} \hat{a}\left(\omega_{j}\right) \exp \left(-i \omega_{j} t_{p}\right),
\end{equation}
which satisfies the canonical commutation relation  $\left[\hat{a}(t_{p}), \hat{a}^{\dag}\left(t_{p’}\right)\right]=\delta_{p, p^{\prime}}$.
We can also obtain the operators in the frequency domain as
\begin{equation}\label{equ3}
\hat{a}\left(\omega_{j}\right)=\frac{1}{\sqrt{j_{B}+1}} \sum^{p_{0}+j_{B}}_{p=p_{0}} \hat{a}\left(t_{p}\right) \exp({i \omega_{j} t_{p}}),
\end{equation}
which forms a pair in the discrete Fourier transformation with $\hat{a}(t_p)$. Furthermore, in the Heisenberg picture, the free evolution of the single frequency mode is $\hat{a}\left(\omega_{j}\right)e^{-i\omega_{j}\tau}$ for the evolution time $\tau$. It suggests that the free evolution of the temporal mode is 
\begin{equation}\label{equ10}
\hat{a}(t_{p}) \rightarrow \hat{a}(t_{p}+\tau)=\frac{1}{\sqrt{j_{B}+1}} \sum_{j=j_{c}-\frac{j_{B}}{2}}^{j_{c}+\frac{j_{B}}{2}} \hat{a}\left(\omega_{j}\right) \exp \left[-i \omega_{j} (t_{p}+\tau)\right].
\end{equation}

\subsection{The process of frequency modulation}

\begin{figure}[h]
\centering
\includegraphics[width=0.7\linewidth]{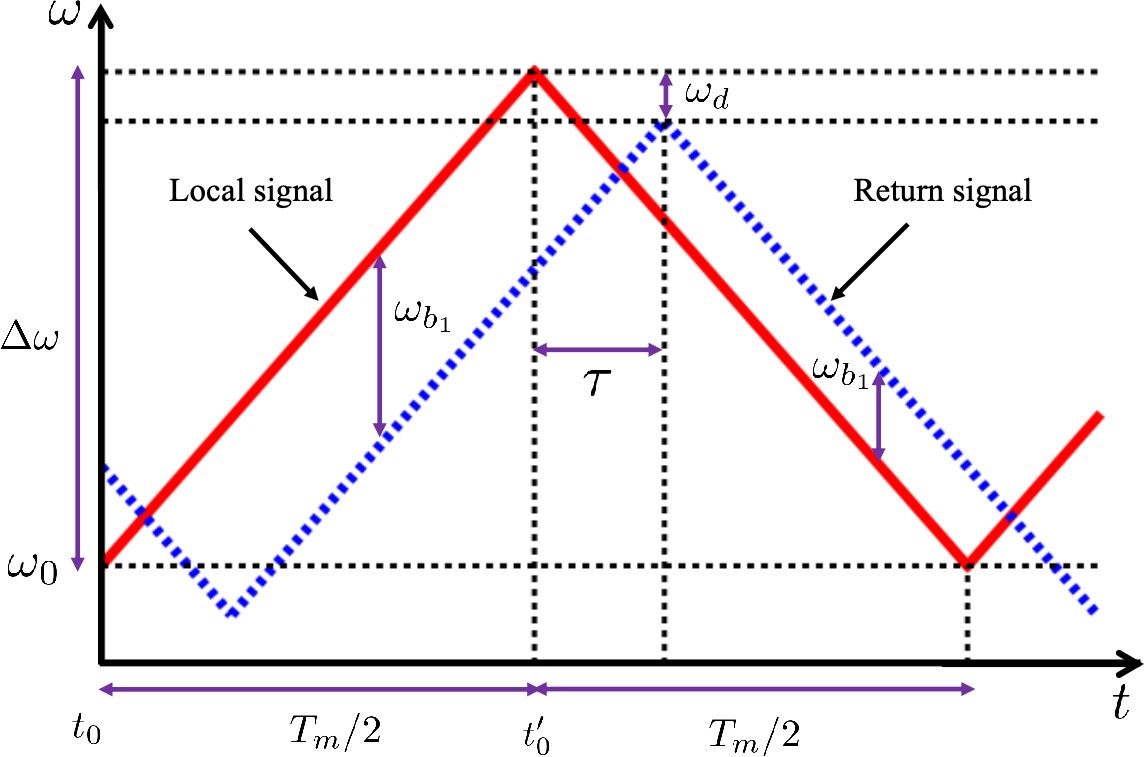}
\caption{The angular frequency $\omega(t)$ of the local and return signals under the triangle frequency modulation with an initial modulation time $t_{0}=t'_{0}-T_{m}/2$.}
\label{fig:supp1}
\end{figure}

Under a finite bandwidth and discrete time approximation, the frequency modulation process can be described by the transformation of the operators in discrete frequency domain as 
\begin{equation}\label{equ8}
\begin{aligned}
\hat{a}_{x}(\omega_{x}) &\rightarrow \sum_{j=j_c-\frac{j_B}{2}}^{j_c+\frac{j_B}{2}} s_{\Delta\omega_{x}}\left(\omega_{j}\right) \hat{a}_{x}\left(\omega_{j}\right),\quad\hat{a}_{x}^{\dag}(\omega_{x}) &\rightarrow \sum_{j=j_c-\frac{j_B}{2}}^{j_c+\frac{j_B}{2}} s_{\Delta\omega_{x}}^{*}\left(\omega_{j}\right) \hat{a}_{x}^{\dag}\left(\omega_{j}\right),
\end{aligned}
\end{equation}
where $\hat{a}_x\left(\omega_x\right)$ and $\hat{a}_x^{\dag}\left(\omega_x\right)$ is the annihilation and generation operator of mode $x$ with frequency $\omega_{x}$, and
\begin{equation}
s_{\Delta\omega_{x}}(\omega_{j})=\frac{1}{T_{m}} \int_{t_{0}}^{t_{0}+T_{m}} E_{\Delta\omega_{x}}(t) e^{-i  \omega_{j} t} d t,
\end{equation}
is the normalized spectrum ($\sum_{j}|s_{\Delta\omega_{x}}(\omega_{j})|^2=1$) of FMCW Laser beam in time domain 
\begin{equation}
E_{\Delta\omega_{x}}(t)=e^{i \phi_{x}\left(t\right)}=\sum_{j} s_{\Delta\omega_{x}}(\omega_{j}) e^{i \omega_{j} t},
\end{equation}
with a time-dependent phase  
\begin{equation}
\phi_{x}\left(t\right)=\int_{t_{0}}^{t} \omega(\Delta\omega_{x},t) d t,
\end{equation}
for initial modulation time $t_{0}$, and modulation period $T_{m}$. For the linear frequency modulation, the angular frequency $\omega(\Delta\omega_{x},t)$ of the FMCW Laser can be expressed as 
\begin{equation}
\omega(\Delta\omega_{x},t)=\frac{\Delta\omega_{x}}{T_{m}}\left(t-t_{0}\right)+\omega_{x},
\end{equation}
when $t\in[t_{0},t_{0}+T_m]$, where $\omega_{x}$ is the initial angular frequency of each period and $\Delta\omega_{x}$ represents the modulation bandwidth of mode $x$. The frequency is linear increase (decrease) for $\Delta\omega_{x}>0$ ($\Delta\omega_{x}<0$).
Further, for the triangular frequency modulation as shown in Fig. \ref{fig:supp1}, the angular frequency $\omega(\Delta\omega_{x},t)$ is a piecewise function
\begin{eqnarray}\label{equ14}
\omega(\Delta\omega_{x},t):=\{\begin{array}{cc}
\frac{\Delta \omega}{T_{m} / 2}\left(t-t'_{0}\right)+(\omega_{x}+\Delta \omega), & \text{for } t\in[t'_{0}-T_{m}/2,t'_{0}],\\ 
-\frac{\Delta \omega}{T_{m} / 2}\left(t-t'_{0}\right)+(\omega_{x}+\Delta \omega), & \text{for } t\in[t'_{0},t'_{0}+T_{m}/2],
\end{array}
\end{eqnarray}
where $t'_{0}$ is the center time of the modulation.

In discrete time domain, the superposition of discrete frequency mode in Eq. (\ref{equ8}) will transform to the discrete temporal mode as
\begin{equation}
 \sum_{j=j_c-\frac{j_B}{2}}^{j_c+\frac{j_B}{2}} s_{\Delta\omega_{x}}\left(\omega_{j}\right) \hat{a}_{x}\left(\omega_{j}\right)\approx\frac{1}{\sqrt{j_B+1}} \sum_{p=p_0}^{p_0+j_B}  E_{\Delta \omega_x}(t_p)\hat{a}_{x}\left(t_p\right),
 \end{equation}
 where Eq. (\ref{equ3}) and the approximation,
\begin{equation}
E_{\Delta \omega_x}(t)\approx\sum_{j=j_c-\frac{j_B}{2}}^{j_c+\frac{j_B}{2}} s_{\Delta \omega_x}\left(\omega_j\right) e^{i \omega_j t},
\end{equation}
is used. Here, we assume that $j_B$ is sufficiently large to cover the majority of the discrete frequency bandwidth. Further, by using Eq. (\ref{equ10}), the free evolution of the mode with frequency modulation is given as
\begin{equation}
\frac{1}{\sqrt{j_B+1}} \sum_{p=p_0}^{p_0+j_B}  E_{\Delta \omega_x}(t_p)\hat{a}_{x}\left(t_p\right)\rightarrow \frac{1}{\sqrt{j_B+1}} \sum_{p=p_0}^{p_0+j_B}  E_{\Delta \omega_x}(t_p-\tau)\hat{a}_{x}\left(t_p\right),
 \end{equation}
assuming the evolution time $\tau \ll T_{m}$.

\subsection{A coherent state with frequency modulation}

Under the discrete time and finite bandwidth approximation, a coherent state ($|\alpha \rangle$) with frequency modulation is presented as \cite{huang2024quantum}
\begin{equation}\label{equ9}
\begin{aligned}
\mathop{\otimes}_{j}\left|\alpha s_{\Delta\omega_{x}}^{*}\left(\omega_{j}\right)\right\rangle 
&\approx\exp \left\{\sum^{j_{c}+\frac{j_{B}}{2}}_{j=j_{c}-\frac{j_{B}}{2}}\left[\alpha s_{\Delta\omega_{x}}^{*}\left(\omega_{j}\right) a^{\dag}\left(\omega_{j}\right)-\alpha^{*} s_{\Delta\omega_{x}}\left(\omega_{j}\right) a\left(\omega_{j}\right)\right]\right\}|0\rangle \\
& =\exp \left\{\frac{1}{\sqrt{j_{B}+1}} \sum^{p_{0}+j_{B}}_{p=p_{0}}\left[\alpha E_{\Delta\omega_{x}}^{*}\left(t_{p}\right) a^{\dag}\left(t_{p}\right)-\alpha^{*} E_{\Delta\omega_{x}}\left(t_{p}\right) a\left(t_{p}\right)\right]\right\}|0\rangle \\
&=\mathop{\otimes}^{p_{0}+j_{B}}_{p=p_{0}}\left|\frac{\alpha E_{\Delta\omega_{x}}^{*}\left(t_{p}\right)}{\sqrt{j_{B}+1}}\right\rangle,
\end{aligned}
\end{equation}
where
\begin{equation}
a\left(t_{p}\right)\left|\frac{\alpha E_{\Delta\omega_{x}}^{*}\left(t_{p}\right)}{\sqrt{j_{B}+1}}\right\rangle=\frac{\alpha E_{\Delta\omega_{x}}^{*}\left(t_{p}\right)}{\sqrt{j_{B}+1}}\left|\frac{\alpha E_{\Delta\omega_{x}}^{*}\left(t_{p}\right)}{\sqrt{j_{B}+1}}\right\rangle,
\end{equation}
is a coherent state in the discrete time domain. As such, the average photon number per unit time for this state is
\begin{equation}
\left\langle\frac{\alpha E_{\Delta\omega_{x}}^*\left(t_p\right)}{\sqrt{j_B+1}}\right| a^{\dag}\left(t_p\right)a\left(t_p\right)\left|\frac{\alpha E_{\Delta\omega_{x}}^*\left(t_p\right)}{\sqrt{j_B+1}}\right\rangle=\frac{|\alpha|^{2}}{j_{B}+1}\equiv n_{coh}.
\end{equation}

\subsection{A two-path-mode squeezed vacuum state with frequency modulation}

A two-path-mode squeezed vacuum state is defined as
\begin{equation}
\hat{S}_2(\xi)|0\rangle=\exp \left[\xi^* \hat{a}_1\left(\omega_1\right) \hat{a}_2\left(\omega_2\right)-\xi \hat{a}_1^{\dag}\left(\omega_1\right) \hat{a}_2^{\dag}\left(\omega_2\right)\right]|0\rangle
\end{equation}
where $\hat{a}_x^{\dag}\left(\omega_x\right)(x=1,2)$ is the generation operator of path mode  $x$ with frequency $\omega_{x}$, and $\xi=r e^{i \theta}$ with squeezing amplitude $r$ and squeezing angle $\theta$. 
By applying the process of frequency modulation as shown in Eq. (\ref{equ8}), a band-limited two-mode squeezed vacuum state with frequency modulation can be represented as
\begin{equation}\label{equ12}
\begin{aligned}
\hat{S}^{FM}_2(\xi)|0\rangle & =\exp \left\{\sum_{m,n=j_c-\frac{j_B}{2}}^{j_c+\frac{j_B}{2}}\left[\xi^* s_{\Delta\omega_1}\left(\omega_m\right) s_{\Delta\omega_2}\left(\omega_n\right) \hat{a}_1\left(\omega_m\right) \hat{a}_2\left(\omega_n\right)-\xi s_{\Delta\omega_1}^*\left(\omega_m\right) s_{\Delta\omega_2}^*\left(\omega_n\right) \hat{a}_1^{\dag}\left(\omega_m\right) \hat{a}_2^{\dag}\left(\omega_n\right)\right]\right\}|0\rangle \\
& \approx\exp \left\{\frac{1}{j_B+1} \sum_{p, q=p_0}^{p_0+j_B}\left[\xi^* E_{\Delta \omega_1}\left(t_p\right) E_{\Delta \omega_2}\left(t_q\right) \hat{a}_1\left(t_p\right) \hat{a}_2\left(t_q\right)-\xi E_{\Delta \omega_1}^*\left(t_p\right) E_{\Delta \omega_2}^*\left(t_q\right) \hat{a}_1^{\dag}\left(t_p\right) \hat{a}_2^{\dag}\left(t_q\right)\right]\right\}|0\rangle \\
& =\exp \left\{\sum_{p, q=p_0}^{p_0+j_B}\left[\xi_{\Delta \omega_1, \Delta \omega_2}^*\left(t_p, t_q\right) \hat{a}_1\left(t_p\right) \hat{a}_2\left(t_q\right)-\xi_{\Delta \omega_1, \Delta \omega_2}\left(t_p, t_q\right) \hat{a}_1^{\dag}\left(t_p\right) \hat{a}_2^{\dag}\left(t_q\right)\right]\right\}|0\rangle
\end{aligned}
\end{equation}
where
\begin{equation}
\xi_{\Delta \omega_1, \Delta \omega_2}\left(t_p, t_q\right) \equiv \frac{1}{j_B+1} \xi E_{\Delta \omega_1}^*\left(t_p\right) E_{\Delta \omega_2}^*\left(t_q\right) = \frac{1}{j_B+1} r e^{i \theta} e^{-i \phi_1\left(t_p\right)} e^{-i \phi_2\left(t_q\right)}.
\end{equation}
Note that, since $\hat{S}_2(\xi)$ and the frequency modulation process both are unitary operation \cite{capmanyQuantumModelElectrooptical2010}, $\hat{S}^{FM}_2(\xi)$ also is a unitary operation.

Further, according to the formula
\begin{equation}
\mathrm{e}^{x \hat{A}} \hat{B} \mathrm{e}^{-x \hat{A}}=B+x[\hat{A}, \hat{B}]+\frac{x^2}{2!}[\hat{A},[\hat{A}, \hat{B}]]+\cdots,
\end{equation}
we find that
\begin{equation}\label{equ4}
\begin{aligned}
&[S_2^{FM}(\xi)]^{\dag} a_1\left(t_{p^{\prime}}\right) \hat{S}^{FM}_2(\xi)\\
&=a_1\left(t_{p^{\prime}}\right)-\sum_{q=p_0}^{p_0+j_B} \xi_{\Delta \omega_1, \Delta \omega_2}\left(t_{p^{\prime}}, t_q\right) a_2^{\dag}\left(t_q\right)
+\frac{1}{2!} \sum_{p, q=p_0}^{p_0+j_B} \xi_{\Delta \omega_1, \Delta \omega_2}^*\left(t_p, t_q\right) \xi_{\Delta \omega_1, \Delta \omega_2}\left(t_{p^{\prime}}, t_q\right) a_1\left(t_p\right)\\
&\quad-\frac{1}{3!} \sum_{p, q, p_1=p_0}^{p_0+j_B} \xi_{\Delta \omega_1, \Delta \omega_2}\left(t_p, t_q\right) \xi^{*}_{\Delta \omega_1, \Delta \omega_2}\left(t_{p_1}, t_q\right) \xi_{\Delta \omega_1, \Delta \omega_2}\left(t_{p_1}, t_{q^{\prime}}\right) a_1^{\dag}\left(t_p\right)+\cdots\\
&=a_1\left(t_{p^{\prime}}\right)-r \frac{1}{j_B+1} e^{i \theta} e^{-i \phi_2\left(t_{q^{\prime}}\right)} \sum_{p=p_0}^{p_0+\jmath_B} e^{-i \phi_1\left(t_p\right)} a_1^{\dag}\left(t_p\right)+\frac{1}{2!} r^2 \frac{1}{j_B+1} e^{-i \phi_2\left(t_{q^{\prime}}\right)} \sum_{q=p_0}^{p_0+j_B} e^{i \phi_2\left(t_q\right)} a_2\left(t_q\right)\\
&\quad-\frac{1}{3!} r^3 \frac{1}{j_B+1} e^{i \theta} e^{-i \phi_2\left(t_{q^{\prime}}\right)} \sum_{p=p_0}^{p_0+j_B} e^{-i \phi_1\left(t_p\right)} a_1^{\dag}\left(t_p\right)+\cdots\\
&=a_1\left(t_{p^{\prime}}\right)+\frac{1}{j_B+1} e^{-i \phi_1\left(t_{p^{\prime}}\right)} \sum_{p=p_0}^{p_0+j_B} e^{i \phi_1\left(t_p\right)} a_1\left(t_p\right)(\cosh r-1)-\frac{1}{j_B+1} e^{i \theta } e^{-i \phi_1\left(t_{p^{\prime}}\right)} \sum_{q=p_0}^{p_0+j_B} e^{-i \phi_2\left(t_q\right)} a_2^{\dag}\left(t_q\right) \sinh r,
\end{aligned}
\end{equation}
and, similarly,
\begin{equation}\label{equ5}
\begin{aligned}
&[\hat{S}^{FM}_2(\xi)]^{\dag} a^{\dag}_1\left(t_{p^{\prime}}\right) \hat{S}^{FM}_2(\xi)\\
&=a^{\dag}_1\left(t_{p^{\prime}}\right)+\frac{1}{j_B+1} e^{i \phi_1\left(t_{p^{\prime}}\right)} \sum_{p=p_0}^{p_0+j_B} e^{-i \phi_1\left(t_p\right)} a^{\dag}_1\left(t_p\right)(\cosh r-1)-\frac{1}{j_B+1} e^{-i \theta } e^{i \phi_1\left(t_{p^{\prime}}\right)} \sum_{q=p_0}^{p_0+j_B} e^{i \phi_2\left(t_q\right)} a_2\left(t_q\right) \sinh r,
\end{aligned}
\end{equation}
\begin{equation}\label{equ6}
\begin{aligned}
&[\hat{S}^{FM}_2(\xi)]^{\dag} a_2\left(t_{q^{\prime}}\right) \hat{S}^{FM}_2(\xi)\\
&=a_2\left(t_{q^{\prime}}\right)+\frac{1}{j_B+1} e^{-i \phi_2\left(t_{q^{\prime}}\right)} \sum_{q=p_0}^{p_0+j_B} e^{i \phi_2\left(t_q\right)} a_2\left(t_q\right)(\cosh r-1)-\frac{1}{j_B+1} e^{i \theta} e^{-i \phi_2\left(t_{q^{\prime}}\right)} \sum_{p=p_0}^{p_0+j_B} e^{-i \phi_1\left(t_p\right)} a_1^{\dag}\left(t_p\right) \sinh r,
\end{aligned}
\end{equation}
\begin{equation}\label{equ7}
\begin{aligned}
&[\hat{S}^{FM}_2(\xi)]^{\dag} a^{\dag}_2\left(t_{q^{\prime}}\right) \hat{S}^{FM}_2(\xi)\\
&=a^{\dag}_2\left(t_{q^{\prime}}\right)+\frac{1}{j_B+1} e^{i \phi_2\left(t_{q^{\prime}}\right)} \sum_{q=p_0}^{p_0+j_B} e^{-i \phi_2\left(t_q\right)} a^{\dag}_2\left(t_q\right)(\cosh r-1)-\frac{1}{j_B+1} e^{-i \theta} e^{i \phi_2\left(t_{q^{\prime}}\right)} \sum_{p=p_0}^{p_0+j_B} e^{i \phi_1\left(t_p\right)} a_1\left(t_p\right) \sinh r.
\end{aligned}
\end{equation}

According to Eqs. (\ref{equ4}-\ref{equ7}), the average photon number per unit time for each path mode of squeezed vacuum state with frequency modulation is
\begin{equation}
\langle 0|[\hat{S}^{FM}_2(\xi)]^{\dag}\hat{a}^{\dag}_{1}\left(t_p\right)\hat{a}_{1}\left(t_p\right)\hat{S}^{FM}_2(\xi)|0\rangle=\langle 0|[\hat{S}^{FM}_2(\xi)]^{\dag}\hat{a}^{\dag}_{2}\left(t_q\right)\hat{a}_{2}\left(t_q\right)\hat{S}^{FM}_2(\xi)|0\rangle=\frac{\sinh^{2}(r)}{j_{B}+1}\equiv n_{sv}.
\end{equation}

\section{Quantum heterodyne detection with sum frequency generation process}

\subsection{Simultaneous measurement of in-phase and quadrature components via quantum heterodyne detection}\label{SMQHD}

\begin{figure}[h]
\centering
\subfigure[]{
    \includegraphics[width=0.4\linewidth]{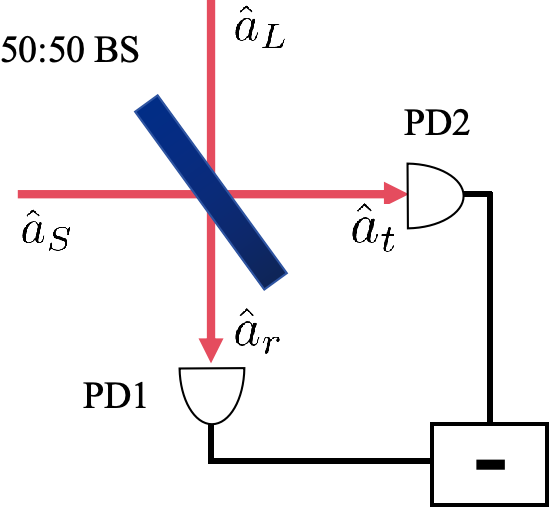}
    \label{fig:subfig1}
}
\subfigure[]{
    \includegraphics[width=0.4\linewidth]{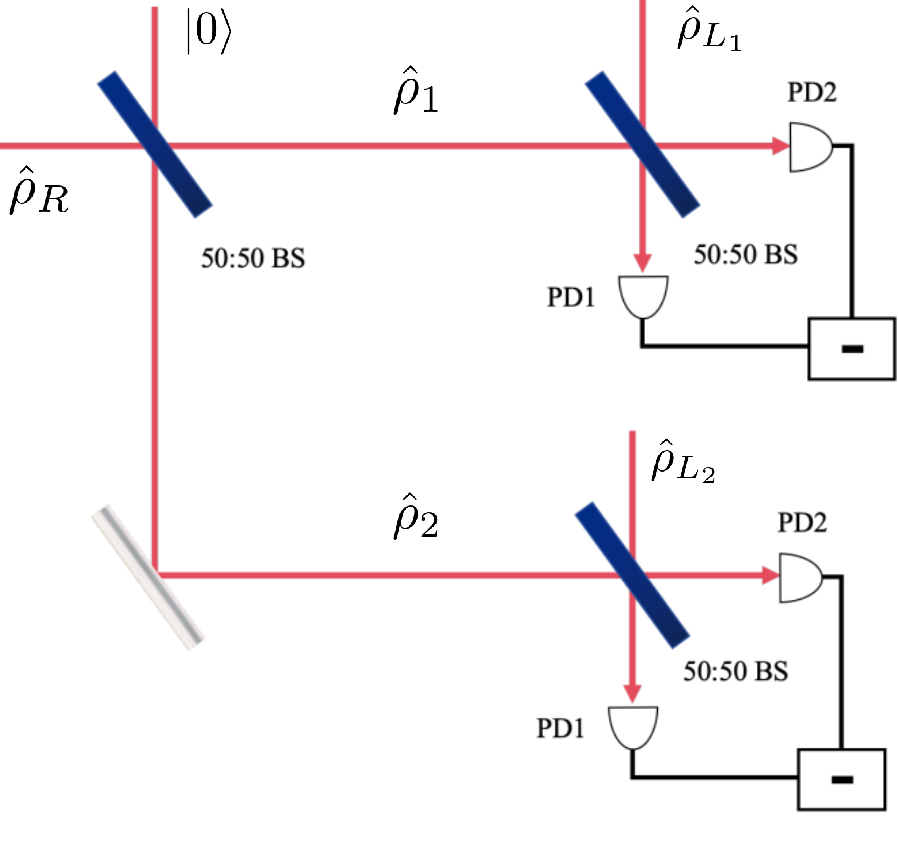}
    \label{fig:subfig2}
}

\caption{The sketch of (a) balanced homodyne detection and (b) one design of phase-diversity homodyne receiver, both employing a 50:50 beam splitter (BS). There is a phase difference of $\pi/2$ between $\hat{\rho}_{L_1}$ and $\hat{\rho}_{L_2}$, which serve as the strong local oscillator modes corresponding to $\hat{\rho}_{1}$ and $\hat{\rho}_{2}$ respectively.}
\label{fig:supp2}
\end{figure}

In the quantum domain, the in-phase and quadrature components of a signal field are extracted using balanced homodyne detection, as illustrated in Fig. \ref{fig:subfig1}, where the signal field $\hat{a}_{S}$ and the local field $\hat{a}_{L}$ share the same frequency and polarization \cite{shapiro2009quantum}. Since the in-phase component $\hat{I}_{S}= (\hat{a}_{S}+\hat{a}^{\dag}_{S})/2$ do not commute with the quadrature component $\hat{Q}_{S}= -i(\hat{a}_{S}-\hat{a}^{\dag}_{S})/2$, it is impossible to measure both components for a given signal state $\rho$ simultaneously with in the setup illustrated in Fig. \ref{fig:subfig1}. To address this challenge, a common approach is to split the state evenly into two paths using a 50:50 beam splitter, resulting in two split states, $\rho_{1}$ and $\rho_{2}$. The in-phase component $\hat{I}_{1}$ of $\rho_{1}$ and the quadrature component $\hat{Q}_{2}$ of $\rho_{2}$ are then measured separately. Since $[\hat{I}_{1}, \hat{Q}_{2}] = 0$ for different paths, $\hat{I}_{1}$ and $\hat{Q}_{2}$ can be measured simultaneously. One design of such a detection system is illustrated in Fig. \ref{fig:subfig2}. This configuration, known as a phase-diversity homodyne receiver, is widely employed in optical communication systems \cite{kikuchi2015fundamentals}. However, splitting the state introduces vacuum noise, which reduces the signal-to-noise ratio (SNR) of the detection \cite{collett1987quantum,kikuchi2015fundamentals}—a trade-off for the ability to measure in-phase and quadrature components simultaneously.

\begin{figure}[h]
\centering
\subfigure[]{
    \includegraphics[width=0.4\linewidth]{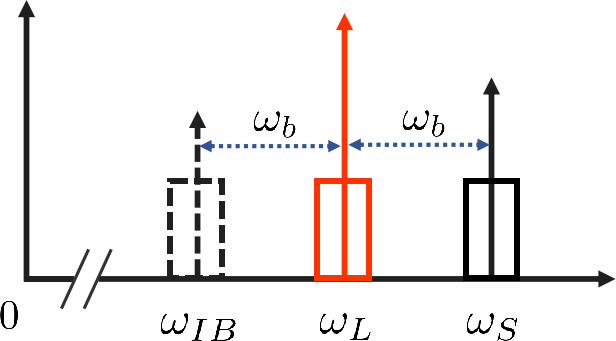}
    \label{fig:supp3_1}
}
\subfigure[]{
    \includegraphics[width=0.4\linewidth]{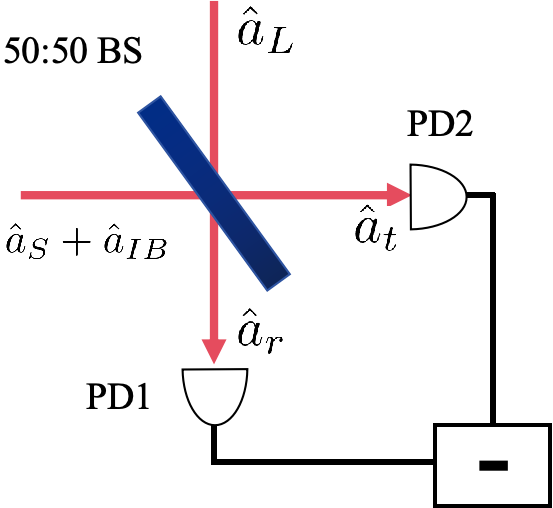}
    \label{fig:supp3_2}
}
\caption{(a) The sketch of signal field, local field, and image-band field spectra, where the frequency of the beat signal $\omega_{b}=\omega_{S}-\omega_{L}=\omega_{L}-\omega_{IB}$. (b) The sketch of balanced heterodyne detection, where 50:50 beam splitter (BS) is being used.}
\label{fig:supp3}
\end{figure}

On the other hand, quantum heterodyne detection (QHD) builds upon the homodyne technique by introducing a frequency difference between the signal field and the local field. In the heterodyne case, the image-band field $\hat{a}_{IB}$, defined as the spectral counterpart of the detected signal field mirrored around the local field frequency (Fig. \ref{fig:supp3_1}), inevitably combines with the signal field, as depicted in Fig. \ref{fig:supp3_2}. The image-band field $\hat{a}_{IB}$ remains in the vacuum state if no signals are injected into this band, which reduces the SNR of the detection \cite{shapiro2009quantum,collett1987quantum,kikuchi2015fundamentals}. Note that, similar to the phase-diversity homodyne receiver, this additional vacuum fluctuation enables the simultaneous measurement of in-phase and quadrature components \cite{kikuchi2015fundamentals}, as described below.

For a multi-frequency-mode CW quantum light field in the discrete time domain, the measurement operator for the intensity difference between the two output ports of a 50:50 beam splitter at $t_{p}$ is represented as
\begin{equation}
\begin{aligned}
n_{d}(t_{p})&=\hat{a}_{r}^{\dag}(t_{p}) \hat{a}_{r}(t_{p})-\hat{a}_{t}^{\dag}(t_{p}) \hat{a}_{t}(t_{p})\\
                  &=\hat{a}_{L}^{\dag}(t_{p}) \hat{a}_{S}(t_{p})+\hat{a}_{S}^{\dag}(t_{p}) \hat{a}_{L}(t_{p})+\hat{a}_{L}^{\dag}(t_{p}) \hat{a}_{IB}(t_{p})+\hat{a}_{IB}^{\dag}(t_{p}) \hat{a}_{L}(t_{p}),
\end{aligned}
\end{equation}
where the input-output relation 
\begin{equation}
\hat{a}_{r}^{\dag}(t_{p})=\frac{1}{\sqrt{2}}[\hat{a}_{S}^{\dag}(t_{p})+\hat{a}_{IB}^{\dag}(t_{p})+\hat{a}_{L}^{\dag}(t_{p})], \quad \hat{a}_{t}^{\dag}(t_{p})=\frac{1}{\sqrt{2}}[\hat{a}_{S}^{\dag}(t_{p})+\hat{a}_{IB}^{\dag}(t_{p})-\hat{a}_{L}^{\dag}(t_{p})],
\end{equation}
is used, and the subscripts $r$, $t$, $S$, $L$, and $IB$ represent the reflection mode, transmission mode, signal mode, local mode, and image-band mode, respectively. The local mode is assumed to be a coherent state with frequency modulation, as described in Eq. (\ref{equ9}), and a sufficiently strong intensity, indicating that
\begin{equation}
\hat{n}_{d}(t_{p})\propto \hat{a}_{S}(t_{p})e^{i \phi_{L}(t_{p})}e^{-i\phi_{0}}+\hat{a}_{S}^{\dag}(t_{p})e^{-i \phi_{L}(t_{p})}e^{i\phi_{0}}+ \hat{a}_{IB}(t_{p})e^{i \phi_{L}(t_{p})}e^{-i\phi_{0}}+\hat{a}_{IB}^{\dag}(t_{p})e^{-i \phi_{L}(t_{p})}e^{i\phi_{0}},
\end{equation}
where $\phi_{0}$ is an additional constant phase given to the local mode. For $\phi_{0}=0$, the intensity difference $\hat{n}_{d}(t_{p})$ is proportional to the sum of two in-phase components:
\begin{equation}\label{equ37}
\hat{n}_{d}(t_{p})\propto \hat{I}_{S}(t_{p})+ \hat{I}_{IB}(t_{p}),
\end{equation}
where the in-phase component for $x=S, IB$ mode in time domain is given as:
\begin{equation}\label{equ11}
\hat{I}_{x}(t_{p})= \frac{1}{2}[\hat{a}_{x}(t_{p})e^{i \phi_{L}(t_{p})}+\hat{a}_{x}^{\dag}(t_{p})e^{-i \phi_{L}(t_{p})}].
\end{equation}
For $\phi_{0}=\pi/2$, the intensity difference $\hat{n}_{d}(t_{p})$ is proportional to the sum of two quadrature components:
\begin{equation}\label{equ38}
\hat{n}_{d}(t_{p})\propto \hat{Q}_{S}(t_{p})+ \hat{Q}_{IB}(t_{p}),
\end{equation}
where the quadrature components for $x=S, IB$ mode in time domain are given as:
\begin{equation}\label{equ36}
 \hat{Q}_{x}(t_{p})= -\frac{i}{2}[\hat{a}_{x}(t_{p})e^{i \phi_{L}(t_{p})}-\hat{a}_{x}^{\dag}(t_{p})e^{-i \phi_{L}(t_{p})}].
\end{equation}
Such that, it is can verified as well,
\begin{equation}\label{equ23}
[\hat{I}_{S}(t_{p}),\hat{Q}_{S}(t_{p})]=[\hat{I}_{IB}(t_{p}),\hat{Q}_{IB}(t_{p})]=\frac{i}{2},
\end{equation}
which is similar to the in-phase and quadrature components in frequency domain. 

If the local mode and the signal mode have a constant beat frequency $\omega_{b} = \omega_{S} - \omega_{L}$, it becomes possible to measure the in-phase and quadrature components of the signal mode simultaneously. However, this measurement comes with additional vacuum noise. To see this, we assure the local mode $\hat{a}_{L}(t_{p})$ is a multi-frequency-mode CW field satisfying
\begin{equation}
\hat{a}_{L}(t_{p}) =\frac{1}{\sqrt{j_{B}+1}} \sum_{j=j_{c}-\frac{j_{B}}{2}}^{j_{c}+\frac{j_{B}}{2}} \hat{a}\left(\omega_{j}\right) \exp \left(-i \omega_{j} t_{p}\right).
\end{equation}
The signal mode $\hat{a}_{S}(t_{p})$ and the image-band mode $\hat{a}_{IB}(t_{p})$ satisfy the following relation as shown in Fig. \ref{fig:supp3_1}, 
\begin{equation}\label{equ39}
\hat{a}_{S}(t_{p}) =\frac{1}{\sqrt{j_{B}+1}} \sum_{j=j_{c}-\frac{j_{B}}{2}}^{j_{c}+\frac{j_{B}}{2}} \hat{a}\left(\omega_{j}+\omega_{b}\right) \exp \left[-i (\omega_{j}+\omega_{b}) t_{p}\right])\equiv \hat{a}^{\prime}_{S}(t_{p})\exp \left(-i \omega_b t_p\right),
\end{equation}
and
\begin{equation}\label{equ40}
\hat{a}_{IB}(t_{p}) =\frac{1}{\sqrt{j_{B}+1}} \sum_{j=j_{c}-\frac{j_{B}}{2}}^{j_{c}+\frac{j_{B}}{2}} \hat{a}\left(\omega_{j}-\omega_{b}\right) \exp \left[-i (\omega_{j}-\omega_{b}) t_{p}\right]\equiv \hat{a}^{\prime}_{IB}(t_{p})\exp \left(i \omega_b t_p\right),
\end{equation}
respectively. By substituting Eq. (\ref{equ39}) and Eq. (\ref{equ40}), the Eq. (\ref{equ37}) can be reformulated as:
\begin{equation}\label{equ41}
\hat{n}_{d}(t_{p})\propto \Big\{\hat{I}^{\prime}_S\left(t_p\right)+\hat{I}^{\prime}_{I B}\left(t_p\right)\Big\}\cos(\omega_{b}t_{p})+\Big\{\hat{Q}^{\prime}_S\left(t_p\right)-\hat{Q}^{\prime}_{I B}\left(t_p\right)\Big\}\sin(\omega_{b}t_{p}),
\end{equation}
where the modified in-phase and quadrature components for the signal mode in time domain are given as:
\begin{equation}
\begin{aligned}
&\hat{I}_S^{\prime}\left(t_p\right)=\frac{1}{2}\left\{\hat{a}_S\left(t_p\right) e^{i\left[\phi_L\left(t_p\right)-\omega_b t_p\right]}+\hat{a}_S^{\dag}\left(t_p\right) e^{-i\left[\phi_L\left(t_p\right)-\omega_b t_p\right]}\right\}, \\
&\hat{Q}^{\prime}_{S}(t_{p})= -i \frac{1}{2}\left\{\hat{a}_{S}(t_{p})e^{i\left[\phi_L\left(t_p\right)-\omega_b t_p\right]}-\hat{a}_{S}^{\dag}(t_{p})e^{-i\left[\phi_L\left(t_p\right)-\omega_b t_p\right]}\right\},
\end{aligned}
\end{equation}
satisfying $[\hat{I}^{\prime}_{S}(t_{p}),\hat{Q}^{\prime}_{S}(t_{p})]=i/2$, and the modified in-phase and quadrature components for the image-band mode in time domain are given as:
\begin{equation}
\begin{aligned}
&\hat{I}_{I B}^{\prime}\left(t_p\right)=\frac{1}{2}\left\{\hat{a}_{I B}\left(t_p\right) e^{i\left[\phi_L\left(t_p\right)+\omega_b t_p\right]}+\hat{a}_{I B}^{\dag}\left(t_p\right) e^{-i\left[\phi_L\left(t_p\right)+\omega_b t_p\right]}\right\}\\
&\hat{Q}_{I B}^{\prime}\left(t_p\right)=-i \frac{1}{2}\left\{\hat{a}_{I B}\left(t_p\right) e^{i\left[\phi_L\left(t_p\right)+\omega_b t_p\right]}-\hat{a}_{I B}^{\dag}\left(t_p\right) e^{-i\left[\phi_L\left(t_p\right)+\omega_b t_p\right]}\right\},
\end{aligned}
\end{equation}
satisfying $[\hat{I}^{\prime}_{IB}(t_{p}),\hat{Q}_{IB}^{\prime}(t_{p})]=i/2$. As such, it is easy to verify that
\begin{equation}
\left[\hat{I}_S^{\prime}\left(t_p\right)+\hat{I}_{I B}^{\prime}\left(t_p\right),\hat{Q}_S^{\prime}\left(t_p\right)-\hat{Q}_{I B}^{\prime}\left(t_p\right)\right]=0,
\end{equation}
which means that the in-phase and quadrature components of the signal mode can be simultaneously measured, albeit with the presence of additional vacuum noise from the image-band mode. Hence, in QHD, measuring the intensity difference $\hat{n}_{d}(t_{p})$ using the setup shown in Fig. \ref{fig:supp3_2} is sufficient to simultaneously determine the in-phase and quadrature components of the signal mode, as described in Eq. (\ref{equ41}).

\subsection{Quantum heterodyne detection following by sum frequency generation process}

In this subsection, we analyze the relationship between the in-phase and quadrature components of the modes before and after the sum-frequency generation (SFG) process. This is particularly relevant as QHD measures the in-phase and quadrature components after the SFG process in one of the FMCW quantum illumination schemes.

A SFG process can be regarded as a time reversed process of spontaneous parametric down-conversion (SPDC). According to Ref. \cite{liu2020joint}, an SFG process in the limit of an infinite SPDC photon bandwidth can be expressed as:
\begin{equation}
\hat{U}_{SFG}=\exp\left\{\epsilon_s\sum^{j_{c}+\frac{j_{B}}{2}}_{m, n=j_{c}-\frac{j_{B}}{2}}[\hat{a}_P^{\dag}\left(\omega_m+\omega_n\right) \hat{a}_S\left(\omega_m\right) a_I\left(\omega_n\right)-\hat{a}_P\left(\omega_m+\omega_n\right) \hat{a}^{\dag}_S\left(\omega_m\right) \hat{a}^{\dag}_I\left(\omega_n\right)]\right\},
\end{equation}
for a multi-frequency-mode CW quantum light field under the finite bandwidth approximation, where $S$, $I$ and $P$ represent signal mode, idler mode, and up-conversion mode (or pump mode for SPDC) respectively. Here, $\epsilon_s$ characterizes the strength of the SFG process. By using Eq. (\ref{equ3}), this SFG process in discrete time domain is given as 
\begin{equation}
\hat{U}_{SFG}=\exp \left\{ \epsilon_{s} \sqrt{j_B+1} \sum_{p=p_0}^{p_0+j_B}[ \hat{a}_P^{\dag}\left(t_p\right) \hat{a}_S\left(t_p\right) \hat{a}_I\left(t_p\right)-\hat{a}_P\left(t_p\right) \hat{a}^{\dag}_S\left(t_p\right) \hat{a}^{\dag}_I\left(t_p\right)]\right\},
\end{equation}
where
\begin{equation}
\begin{aligned}
& \sum_{m, n} \hat{a}_P^{\dag}\left(\omega_m+\omega_n\right) \hat{a}_S\left(\omega_m\right) a_I\left(\omega_n\right) \\
= & \left(\frac{1}{\sqrt{j_B+1}}\right)^3 \sum_{p, q, q^{\prime}=p_0}^{p_0+j_B} \hat{a}_P^{\dagger}\left(t_p\right) \hat{a}_S\left(t_q\right) \hat{a}_I\left(t_{q^{\prime}}\right) \sum_{m, n} \exp \left[-i \omega_m\left(t_p-t_q\right)\right] \exp \left[-i \omega_n\left(t_p-t_{q^{\prime}}\right)\right] \\
= & \sqrt{j_B+1} \sum_{p=p_0}^{p_0+j_B} \hat{a}_P^{\dag}\left(t_p\right) \hat{a}_S\left(t_p\right) \hat{a}_S\left(t_p\right).
\end{aligned}
\end{equation}
Further, when the mean photon number in the idler mode is sufficiently small, the SFG process can be approximated as 
\begin{equation}\label{equ42}
\hat{U}_{S F G}\approx I_{s}+\epsilon_s \sqrt{j_B+1}\sum_{p=p_{0}}^{p_0+j_B}\left[\hat{a}_P^{\dag}\left(t_p\right) \hat{a}_S\left(t_p\right) \hat{a}_I\left(t_p\right)-\hat{a}_P\left(t_p\right) \hat{a}_S^{\dag}\left(t_p\right) \hat{a}_I^{\dag}\left(t_p\right)\right].
\end{equation}
where Taylor expansion is used. Here, $I_{s}$ represents an identity operation that does not generate any up-converted modes. In practice, the unconverted mode can be filtered out using a high-pass filter.

Since the image-band mode discussed in last subsection will not participate in SFG process, we only consider the relationship between the in-phase and quadrature components of the up-converted before and after the SFG process. According to Eq. (\ref{equ42}), the equivalent in-phase and quadrature components of up-conversion mode before the SFG process are
\begin{equation}\label{equ19}
\hat{U}_{SFG}^{\dag} \hat{I}_P(t_{p}) \hat{U}_{SFG} \propto \epsilon_s \frac{1}{2}\left[\hat{a}_S\left(t_p\right) \hat{a}_I\left(t_p\right) e^{i \phi_L(t_p)}+\hat{a}_S^{\dag}\left(t_p\right) \hat{a}_I^{\dag}\left(t_p\right) e^{-i \phi_L(t_p)}\right],
\end{equation}
and
\begin{equation}\label{equ20}
\hat{U}_{SFG}^{\dag} \hat{Q}_P(t_{p}) \hat{U}_{SFG} \propto -\epsilon_s \frac{i}{2}\left[\hat{a}_S\left(t_p\right) \hat{a}_I\left(t_p\right) e^{i \phi_L(t_p)}-\hat{a}_S^{\dag}\left(t_p\right) \hat{a}_I^{\dag}\left(t_p\right) e^{-i \phi_L(t_p)}\right],
\end{equation}
if $\epsilon_s\neq 0$. Here, the $P$ mode on the right side is omitted since it does not have a pump mode for the SFG process. Further, by using Eq. (\ref{equ11}) and Eq. (\ref{equ36}), it is easy to find that
\begin{equation}
\hat{a}_S\left(t_p\right)=\left[\hat{I}_S\left(t_p\right)+i \hat{Q}_S\left(t_p\right)\right] e^{i \phi_{L_{S}}\left(t_p\right)},\quad \hat{a}_I\left(t_p\right)=\left[\hat{I}_I\left(t_p\right)+i \hat{Q}_I\left(t_p\right)\right] e^{i \phi_{L_{I}}\left(t_p\right)},
\end{equation}
where $L_{S}$ and $L_{I}$ is the local mode for signal and idler mode respectively.
It suggests that Eq. (\ref{equ19}) and Eq. (\ref{equ20}) can be represented as follows
\begin{equation}\label{equ21}
\hat{U}_{SFG}^{\dag} \hat{I}_P(t_{p}) \hat{U}_{SFG} \propto \epsilon_s\left[\hat{I}_S\left(t_p\right) \hat{I}_I\left(t_p\right)-\hat{Q}_S\left(t_p\right) \hat{Q}_I\left(t_p\right)\right],
\end{equation}
and
\begin{equation}\label{equ22}
\hat{U}_{SFG}^{\dag} \hat{Q}_P(t_{p}) \hat{U}_{SFG} \propto \epsilon_s\left[\hat{I}_S\left(t_p\right) \hat{Q}_I\left(t_p\right)+\hat{Q}_S\left(t_p\right) \hat{I}_I\left(t_p\right)\right]
\end{equation}
where $\phi_L(t_p)=\phi_{L_{S}}(t_p)+\phi_{L_{I}}(t_p)$. The Eq. (\ref{equ21}) and Eq. (\ref{equ22}) relate the in-phase and quadrature components before and after the SFG process. In this case, we define 
\begin{equation}
 \hat{I}^{\prime}_P(t_{p})  \equiv \hat{I}_S\left(t_p\right) \hat{I}_I\left(t_p\right)-\hat{Q}_S\left(t_p\right) \hat{Q}_I\left(t_p\right), \quad  \hat{Q}^{\prime}_P(t_{p})  \equiv \hat{I}_S\left(t_p\right) \hat{Q}_I\left(t_p\right)+\hat{Q}_S\left(t_p\right) \hat{I}_I\left(t_p\right),
\end{equation}
as equivalent in-phase and quadrature components of up-conversion mode. For a quantum optical state $\rho$ with signal and idler mode before the SFG process,
\begin{equation}
\operatorname{Tr}\left(\hat{\rho} [\hat{I}^{\prime}_{P}(t_{p}),\hat{Q}^{\prime}_{P}(t_{p})]\right)=\frac{i}{2}(1+n_{S}+n_{I}),
\end{equation}
where $n_{S}$ and $n_{I}$ are the mean photon number of the signal and idler mode in per unit time, respectively. As such, the commutation relation $[\hat{I}^{\prime}_{P}(t_{p}),\hat{Q}^{\prime}_{P}(t_{p})]$ will reduce to Eq. (\ref{equ23}) when the mean photon number of per temporal mode in both the signal and idler modes is sufficiently small, i.e., $n_{S}\ll1$ and $n_{I}\ll1$.

\section{The quantum limit of estimating beat information for Gaussian states}

\subsection{The Wigner function of Gaussian states with beat information}

A Gaussian state \cite{weedbrook2012gaussian} is completely characterized by its Wigner function, which follows a Gaussian distribution. This implies that the Wigner function of a Gaussian state can be fully determined from the mean values and covariance matrix of the in-phase and quadrature components. Examples of common Gaussian states include the vacuum state, coherent state, thermal state, squeezed state, and two-mode squeezed state.

Specifically, for $m$-mode Gaussian state $\hat{\rho}$, the Wigner function is defined as \cite{liuQuantumFisherInformation2020}
\begin{equation}\label{equ29}
W(\vec{X})=\frac{1}{(2 \pi)^m \sqrt{\operatorname{Det} C}} \mathrm{e}^{-\frac{1}{2}(\vec{X}-\langle\vec{X}\rangle)^{\mathrm{T}} C^{-1}(\vec{X}-\langle\vec{X}\rangle)},
\end{equation}
where $\vec{X}=\left(\hat{I}_1, \hat{Q}_1, \ldots, \hat{I}_m, \hat{Q}_m\right)^{\mathrm{T}}$ can be regarded as the operators of generalized coordinates and generalized momentum of the $m$-particle system, which has mean values $\langle\vec{R}\rangle_j=\operatorname{Tr}\left(\hat{X}_j \hat{\rho}\right)$ and covariance matrix
\begin{equation}
C_{i j}:=\operatorname{cov}_\rho\left(\hat{X}_i, \hat{X}_j\right)=\frac{1}{2} \operatorname{Tr}\left(\hat{\rho}\left\{\hat{X}_i, \hat{X}_j\right\}\right)-\operatorname{Tr}\left(\hat{\rho} \hat{X}_i\right) \operatorname{Tr}\left(\hat{\rho} \hat{X}_j\right),
\end{equation}
with anticommutation relation $\{\hat{X}_i, \hat{X}_j\}=\hat{X}_i\hat{X}_j+\hat{X}_j\hat{X}_i$. For a multi-frequency-mode Gaussian state, the in-phase and quadrature components in the time domain, as described in Eq. (\ref{equ11}), are used to define the generalized coordinates and generalized momentum operators, $\vec{X}$, within the Wigner function. In this case, the Wigner function may depend on the beat frequency, which is determined by the frequency difference between the Gaussian state and the local oscillator light. To analyze the beat frequency information encoded in the Wigner function, a single temporal mode of the Gaussian state for each particle is considered, as shown below.

After experiencing a free evolution and a Doppler shift, a single temporal mode of the coherent state with frequency modulation is
\begin{equation}\label{equ16}
\left|\frac{\alpha E_{\Delta \omega_S}^*\left(t_p-\tau,\omega_{S}-\omega_{d}\right)}{\sqrt{j_B+1}}\right\rangle\equiv|\alpha’ E^{*}_{\Delta \omega_S}{(t_p-\tau,\omega_{S}-\omega_{d})}\rangle,
\end{equation}
where $\alpha’\equiv\alpha/\sqrt{j_{B}+1}$ and $\omega_{d}$ is the Doppler shift. If apply the same triangular frequency modulation to the signal and local light as shown in Fig. \ref{fig:supp1}, the mean values of in-phase and quadrature components in the time domain are given as
\begin{equation}
\left\langle I_S(t_{p})\right\rangle_{coh}=\alpha_1^{\prime} \cos \left(\omega_l t_{p}+\phi_l\right),\quad \left\langle Q_S(t_{p})\right\rangle_{coh}=\alpha_1^{\prime} \sin \left(\omega_l t_{p}+\phi_l\right),
\end{equation}
where 
\begin{equation}\label{equ15}
\omega_l=- \frac{\Delta \omega}{T_m / 2} \tau+(-1)^l\omega_{d},\quad \theta_l=(-1)^l[\left(\omega_0-\omega_d +\Delta \omega\right) \tau-\omega_{l} t_{d_l}],
\end{equation}
in which $l=1$ for $t \in\left[t_{d_0}-T_m / 2, t_{d_0}\right]$ with a initial detection time $t_{d_0}-T_m / 2$,  $l=0$ for $t \in\left[t_{d_1}, t_{d_1}+T_m / 2\right]$ with a initial detection time $t_{d_1}$, and assuming $T_{m}\gg\tau$. Further, the covariance matrix of of in-phase and quadrature components in the time domain is given as
\begin{equation}
C_{coh}=\frac{1}{4}\left(\begin{array}{ll}
1 &\quad 0 \\
0 &\quad 1
\end{array}\right),
\end{equation}
which is independent of time. 

For the two-path-mode squeezed vacuum state with frequency modulation defined in Eq. (\ref{equ12}), separating a single temporal mode for each path is challenging, as the state cannot be expressed as a product state in the time domain. However, using Eq. (\ref{equ4}) through Eq. (\ref{equ7}), the mean values and covariance matrix of the in-phase and quadrature components for a single temporal mode of each path in the time domain can be effectively calculated. For simplicity, we define the path mode 1 as signal mode $S$ and the path mode 2 as idler mode $I$, and only the signal mode experiences the free evolution and the Doppler shift. In this case, the two-path-mode squeezed vacuum state with frequency modulation becomes
\begin{equation}\label{equ18}
\begin{aligned}
&\hat{S}^{FM}_2(\xi,\tau,\omega_d)|0\rangle\\
 &=\exp \left\{\sum_{p, q=p_0}^{p_0+j_B}\left[\xi_{\Delta \omega_S, \Delta \omega_I}^*\left(t_p-\tau, \omega_S-\omega_d, t_q,\omega_I\right) \hat{a}_S\left(t_p\right) \hat{a}_I\left(t_q\right)-\xi_{\Delta \omega_S, \Delta \omega_I}\left(t_p-\tau,\omega_S-\omega_d, t_q, \omega_I\right)\hat{a}_S^{\dag}\left(t_p\right) \hat{a}_I^{\dag}\left(t_q\right)\right]\right\}|0\rangle
\end{aligned}
\end{equation}
where
\begin{equation}
\xi_{\Delta \omega_1, \Delta \omega_2}\left(t_p-\tau,  \omega_S-\omega_d, t_q,\omega_I\right) = \frac{1}{j_B+1} r e^{i \theta} e^{-i \phi_S\left(t_p-\tau,\omega_S-\omega_d\right)} e^{-i \phi_I\left(t_q,\omega_I\right)},
\end{equation}
and $\xi=r e^{i \theta}$. Then, assuming the local light in signal and idler mode have the same frequency at the same time, the covariance between the in-phase components $\hat{I}_{S}(t_{p’})$ and $\hat{I}_{I}(t_{p’})$ in a single temporal mode is given as
\begin{equation}
\begin{aligned}
\operatorname{cov}\left(\hat{I}_{S}(t_{p’}), \hat{I}_{I}(t_{p’})\right)&=\langle 0|[\hat{S}^{FM}_2(\xi,\tau,\omega_d)]^{\dag}\hat{I}_{S}(t_{p’})\hat{I}_{I}(t_{p’})\hat{S}^{FM}_2(\xi,\tau,\omega_d)|0\rangle\\
&=\frac{1}{4}\langle 0|[\hat{S}_2^{F M}\left(\xi, \tau, \omega_d\right)]^{\dag}\hat{a}^{\dag}_{S}(t_{p’})\hat{a}_{I}(t_{p’})\hat{S}_2^{F M}\left(\xi, \tau, \omega_d\right)|0\rangle \\
&+\frac{1}{4}\langle 0|[\hat{S}_2^{F M}\left(\xi, \tau, \omega_d\right)]^{\dag}\hat{a}_{S}(t_{p’})\hat{a}^{\dag}_{I}(t_{p’})\hat{S}_2^{F M}\left(\xi, \tau, \omega_d\right)|0\rangle \\
&+\frac{1}{4}\langle 0|[\hat{S}_2^{F M}\left(\xi, \tau, \omega_d\right)]^{\dag}\hat{a}_{S}(t_{p’})\hat{a}_{I}(t_{p’})\hat{S}_2^{F M}\left(\xi, \tau, \omega_d\right)|0\rangle e^{2i \phi_{L}(t_{p’})}\\
&+\frac{1}{4}\langle 0|[\hat{S}_2^{F M}\left(\xi, \tau, \omega_d\right)]^{\dag}\hat{a}^{\dag}_{S}(t_{p’})\hat{a}^{\dag}_{I}(t_{p’})\hat{S}_2^{F M}\left(\xi, \tau, \omega_d\right)|0\rangle e^{-2i \phi_{L}(t_{p’})}
\end{aligned}
\end{equation}
where the following conditions
\begin{equation}
\langle 0|[\hat{S}^{FM}_2(\xi,\tau,\omega_d)]^{\dag}\hat{I}_{S}(t_{p})\hat{S}^{FM}_2(\xi,\tau,\omega_d)|0\rangle=0, \quad \langle 0|[\hat{S}^{FM}_2(\xi,\tau,\omega_d)]^{\dag}\hat{I}_{I}(t_{p})\hat{S}^{FM}_2(\xi,\tau,\omega_d)|0\rangle=0,
\end{equation}
and $[\hat{I}_{S}(t_{p}),\hat{I}_{I}(t_{p})]=0$ are considered. Further, using Eqs. (\ref{equ4}-\ref{equ7}) and the unitary property of $\hat{S}^{FM}_2(\xi,\tau,\omega_d)$, this covariance is calculated as 
\begin{equation}\label{equ13}
\operatorname{cov}\left(\hat{I}_{S}(t_{p’}), \hat{I}_{I}(t_{p’})\right)=-\frac{\sqrt{n_{sv}(n_{sv}+1)}}{2} \cos\left[2\phi_L\left(t_{p^{\prime}}\right)-\phi_S\left(t_p-\tau, \omega_S-\omega_d\right)-\phi_I\left(t_q,\omega_I\right)\right]
\end{equation}
where $n_{sv}=\sinh^{2}(r)$ assuming there are $j_{B}+1$ identical copies of the two-path-mode squeezed vacuum state with frequency modulation at per unit time, and $\theta=0$ for simplicity. 

According to Eq. (\ref{equ13}), for a local mode with a fixed frequency, the covariance exhibits single-tone beating when the sum of the frequencies of the signal and idler modes remains constant. Therefore, if the triangular frequency modulation $\omega(\Delta\omega_{S},t)$ from Eq. (\ref{equ14}) is applied to the signal mode, the frequency modulation $\omega(-\Delta\omega_{I},t)$ should be applied to the idler mode. This means that when the frequency of the signal mode increases (or decreases) linearly, the frequency of the idler mode decreases (or increases) linearly with the same slope. In this case, setting the frequency of local mode $\omega_{L}=\omega_{S}=\omega_{I}$, the covariance Eq. (\ref{equ13}) becomes
\begin{equation}
\operatorname{cov}\left(\hat{I}_{S}(t_{p’}), \hat{I}_{I}(t_{p’})\right)=-\frac{\sqrt{n_{sv}(n_{sv}+1)} }{2}\cos\left(\omega_l t_p+\phi_l\right),
\end{equation}
 where $\omega_l$ and $\phi_l$ is the same as Eq. (\ref{equ15}). Further, by calculating the covariances individually, the covariance matrix that contains the beat frequency information is given as
\begin{equation}
C_{sv}=\frac{1}{4}\left(\begin{array}{ll}
\quad\quad\quad(1+2 n_{sv})I_{2} & 2 \sqrt{n_{s v}\left(1+n_{s v}\right)}\Lambda(\omega_{l},\phi_l) \\
2 \sqrt{n_{s v}\left(1+n_{s v}\right)}\Lambda(\omega_{l},\phi_l) & \quad\quad\quad(1+2 n_{sv})I_{2}
\end{array}\right),
\end{equation}
where
\begin{equation}\label{equ35}
\Lambda\left(\omega_l, \phi_l\right)=\left(\begin{array}{ll}
-\cos \left(\omega_l t_p+\phi_l\right) & -\sin \left(\omega_l t_p+\phi_l\right) \\
-\sin \left(\omega_l t_p+\phi_l\right) & \cos \left(\omega_l t_p+\phi_l\right)
\end{array}\right), \quad I_{2}=\left(\begin{array}{ll}
1 & 0 \\
0& 1
\end{array}\right).
\end{equation}

\subsection{The Gaussian noise and operation}

Some Gaussian states are independent from the beat frequency which can be considered as the additional noise of the system. For example, the vacuum state $|0\rangle$ has the same covariance matrix as the coherent state
\begin{equation}
C_{vac}=\frac{1}{4}\left(\begin{array}{ll}
1 &\quad 0 \\
0 &\quad 1
\end{array}\right),
\end{equation}
but with zero mean value, that is,  
\begin{equation}
\left\langle I_S(t_{p})\right\rangle_{vac}=0,\quad \left\langle Q_S(t_{p})\right\rangle_{vac}=0.
\end{equation}

Another common type of quantum noise is thermal noise. For a multi-frequency-mode continuous wave, thermal state with a finite bandwidth has the following form 
\begin{equation}\label{equ17}
\rho_{th}=\mathop{\otimes}^{j_c+\frac{j_B}{2}}_{j=j_c-\frac{j_B}{2}}\rho_{th}(\omega_{j})=\mathop{\otimes}^{j_c+\frac{j_B}{2}}_{j=j_c-\frac{j_B}{2}}\int \mathrm{d}^2 \alpha(\omega_{j})|\alpha(\omega_{j})\rangle\langle\alpha(\omega_{j})| P[\alpha(\omega_{j})]
\end{equation}
which is the product of thermal state for each frequency mode. For each frequency mode, this state is a mixture of coherent state with coefficient
\begin{equation}
P\left[\alpha\left(\omega_j\right)\right]=\frac{1}{\pi \bar{n}_j} \mathrm{e}^{-\left|\alpha\left(\omega_j\right)\right|^2 / \bar{n}_j}
\end{equation}
where $\bar{n}_j$ is the mean photon number of the thermal state with a frequency of 
$\omega_{j}$. For this state, the covariance matrix of in-phase and quadrature components in the time domain are given as
\begin{equation}
C_{th}=\frac{1}{4}\left(\begin{array}{ll}
1+2n_{th} &\quad \quad 0 \\
\quad 0 &\quad 1+2n_{th}
\end{array}\right)
\end{equation}
where 
\begin{equation}
n_{th}=\sum^{j_c+\frac{j_B}{2}}_{j=j_c-\frac{j_B}{2}}n_{j}/(j_{B}+1) 
\end{equation}
is the mean photon number per mode. Furthermore, the thermal state has zero mean for both the in-phase and quadrature components, similar to the vacuum state:
\begin{equation}
\left\langle I_S(t_{p})\right\rangle_{th}=0,\quad \left\langle Q_S(t_{p})\right\rangle_{th}=0.
\end{equation}
As such, for $n_{th}=0$, the thermal state reduces to the vacuum state.

The Gaussian operation \cite{weedbrook2012gaussian}, on the other hand, is a unitary operation that maps a Gaussian state to another Gaussian state. The only Gaussian operation applied afterward is an asymmetric beam splitter acting on a two-particle Gaussian state
\begin{equation}
BS(\epsilon)=\left(\begin{array}{cccc}
\sqrt{\epsilon}\quad & 0 \quad& -\sqrt{1-\epsilon} \quad& 0 \\
0 \quad& \sqrt{\epsilon} \quad& 0 \quad& -\sqrt{1-\epsilon} \\
\sqrt{1-\epsilon} \quad& 0 \quad& \sqrt{\epsilon} \quad& 0 \\
0 \quad& \sqrt{1-\epsilon} \quad& 0 \quad& \sqrt{\epsilon}
\end{array}\right),
\end{equation}
    where $\epsilon$ is the reflectivity of the beam splitter (BS), which suggests an input-output relationship that is given as
\begin{equation}
\left(\begin{array}{l}
\hat{I}_1^{in}\left(t_p\right) \\
\hat{Q}_1^{in}\left(t_p\right) \\
\hat{I}_2^{in}\left(t_p\right) \\
\hat{Q}_2^{in}\left(t_p\right)
\end{array}\right)=BS^{-1}(\epsilon)\left(\begin{array}{l}
\hat{I}_1^{out}\left(t_p\right) \\
\hat{Q}_1^{out}\left(t_p\right) \\
\hat{I}_2^{out}\left(t_p\right) \\
\hat{Q}_2^{out}\left(t_p\right)
\end{array}\right)=\left(\begin{array}{l}
\sqrt{\epsilon}\hat{I}^{out}_{1}(t_{p}) + \sqrt{1-\epsilon} \hat{I}^{out}_{2}(t_{p})  \\
\sqrt{\epsilon}\hat{Q}^{out}_{1}(t_{p}) + \sqrt{1-\epsilon} \hat{Q}^{out}_{2}(t_{p}) \\
- \sqrt{1-\epsilon}\hat{I}^{out}_{1}(t_{p})+\sqrt{\epsilon}\hat{I}^{out}_{2}(t_{p})  \\
-\sqrt{1-\epsilon}\hat{Q}^{out}_{1}(t_{p}) +\sqrt{\epsilon}\hat{Q}^{out}_{2}(t_{p}) 
\end{array}\right).
\end{equation}
When the signal light field is simultaneously affected by losses and thermal noise, this can be modeled as mixing the signal light field with a thermal light field on an asymmetric beam splitter, while retaining only the modes along the signal path.

\subsection{The quantum Cramér-Rao bound of the Gaussian state with beat information}

In parameter estimation theory, the quantum Cramér-Rao bound (CRB) establishes a lower bound for the variance of an unknown parameter $x$ within a quantum state $\rho(x)$
\begin{equation}
E[(\hat{x}-x)^{2}]=\delta^{2} \hat{x} \geqslant \frac{1}{\nu } F^{-1} _{Q}(x),
\end{equation}
where $\nu$ represents the number of times the procedure is repeated, $\hat{x}$ is the unbiased estimators of a $x$, $F_{Q}(x)$ is the quantum Fisher information (QFI) which only depends on the quantum state $\rho(x)$. As shown in Eq. (\ref{equ15}), both the frequency and the initial phase of the beat are related to the time of flight $\tau$ and the Doppler shift $\omega_d$. Due to the $2\pi$ ambiguity in measuring phase, the beat frequency is used to estimate $\tau$ and $\omega_d$ simultaneously. Then, the unknown parameter to be estimated is $x=\omega_{l}$.

For the frequency-modulated coherent state shown in Eq. (\ref{equ16}) mixed with a multi-frequency-mode CW thermal state shown in Eq. (\ref{equ17}), the mean values and the covariance matrix of in-phase and quadrature components in the time domain of the return state $\rho_{coh-th}$ are given as
\begin{equation}\label{equ30}
\left\langle I_R(t_{p})\right\rangle_{coh-th}=\sqrt{\epsilon}\alpha^{\prime} \cos \left(\omega_l t_{p}+\phi_l\right),\quad \left\langle Q_R(t_{p})\right\rangle_{coh-th}=\sqrt{\epsilon}\alpha^{\prime} \sin \left(\omega_l t_{p}+\phi_l\right),
\end{equation}
where the subscript $R$ represents the return mode, and
\begin{equation}\label{equ31}
C_{coh-th}=\epsilon C_{coh}+(1-\epsilon)C_{th}.
\end{equation}
Here, $\epsilon$ represents the reflectivity of the beam splitter used to mix the two states, which can also be interpreted as the reflectivity of the detected target. Then, using the method described in the literature \cite{nichols2018multiparameter} for computing the QFI of Gaussian states, the QFI of this mixed state per temporal mode is given as
\begin{equation}\label{equ26}
F_{coh-th}^{Q}(\omega_l,t_{p})=\frac{4 t_{p}^2 n_{coh} \epsilon}{1+2 n_{t h}(1-\epsilon)}\approx\frac{4 t^2 n_{coh} \epsilon}{1+2 n_{t_{p} h}},
\end{equation}
where the approximation holds for $\epsilon\ll 1$.

For the two-path-mode squeezed vacuum state with frequency modulation shown in Eq. (\ref{equ18}) mixed with a multi-frequency-mode CW thermal state shown in Eq. (\ref{equ17}), the mean values and the covariance matrix of in-phase and quadrature components in the time domain of the return state $\rho_{sv-th}$ are given as
\begin{equation}
\left\langle I_R(t_{p})\right\rangle_{sv-th}=0,\quad \left\langle Q_R(t_{p})\right\rangle_{sv-th}=0, \quad \left\langle I_I(t_{p})\right\rangle_{sv-th}=0,\quad \left\langle Q_I(t_{p})\right\rangle_{sv-th}=0,
\end{equation}
and
\begin{equation}\label{equ33}
C_{sv-th}=\frac{1}{4}\left(\begin{array}{cc}
\epsilon(1+2 n_{sv})I_{2}+(1-\epsilon)(1+2 n_{th})I_{2} &\quad2 \sqrt{\epsilon n_{s v}\left(1+n_{s v}\right)}\Lambda(\omega_{l},\phi_l) \\
 2 \sqrt{\epsilon n_{s v}\left(1+n_{s v}\right)}\Lambda(\omega_{l},\phi_l) &\quad (1+2 n_{sv})I_{2}
\end{array}\right),
\end{equation}
where only the signal mode experiences the loss and thermal noise. Then, the QFI of this mixed state per temporal mode is given as
\begin{equation}\label{equ27}
F_{sv-th}^{Q}(\omega_l,t_{p})=\frac{4 n_{s v}\left(1+n_{s v}\right) t_{p}^2 \epsilon}{1+n_{s v}(1-\epsilon)+n_{t h}\left(1+2 n_{s v}\right)(1-\epsilon)}\approx\frac{4 t_{p}^2 n_{sv}\epsilon}{1+n_{th}},
\end{equation}
where the approximation holds for $\epsilon \ll 1$ and $n_{s v} \ll 1$. When $n_{coh} = n_{sv} $, Eqs. (\ref{equ26}) and (\ref{equ27}) indicate that, in the limit $\epsilon \ll 1 $ and  $n_{sv} \ll 1$, the QFI for both the coherent state and the two-path-mode squeezed vacuum state is the same, i.e., $F_{sv-th}^{Q}(\omega_l,t_{p}) = F_{coh-th}^{Q}(\omega_l,t_{p})$, under weak thermal noise $ n_{th} \ll 1 $. However, in the case of strong thermal noise $ n_{th} \gg 1 $, the QFI of the two-path-mode squeezed vacuum state is twice that of the coherent state, $ F_{sv-th}^{Q}(\omega_l,t_{p}) = 2F_{coh-th}^{Q}(\omega_l,t_{p}) $, indicating a 3 dB improvement.

If the signal and idler light of the mixed state $\rho_{sv-th}$ undergo an additional sum-frequency generation (SFG) process, in the limit of $\epsilon \ll 1$ and $n_{s v} \ll 1$, the mean values and the covariance matrix of the in-phase and quadrature components of the up-converted state $\rho_{SFG}$ are given as follows
\begin{equation}\label{equ24}
\left\langle I^{\prime}_P(t_{p})\right\rangle_{SFG}=-\sqrt{\epsilon n_{s v}} \cos \left(\omega_l t_{p}+\phi_l\right),\quad \left\langle Q^{\prime}_P(t_{p})\right\rangle_{SFG}=-\sqrt{\epsilon n_{s v}} \sin \left(\omega_l t_{p}+\phi_l\right),
\end{equation}
and
\begin{equation}\label{equ25}
C^{\prime}_{SFG}=\frac{1}{8}\left(\begin{array}{cc}
1+2 n_{t h}+4 n_{sv} \epsilon \cos(2\omega_l t_{p}+2\phi_l) & 4 n_{sv} \epsilon \sin(2\omega_l t_{p}+2\phi_l) \\
4 n_{sv} \epsilon \sin(2\omega_l t_{p}+2\phi_l)  & 1+2 n_{t h}-4 n_{sv} \epsilon \cos(2\omega_l t_{p}+2\phi_l)]
\end{array}\right).
\end{equation}
Note that, although both the mean values and the covariance matrix of this up-converted mode contain information about the beat, the contribution of the covariance matrix to the QFI can be neglected when $n_{sv} \epsilon \ll 1$. To illustrate this, the QFI of this mixed state, excluding the contribution of the covariance matrix, is given as follows:
\begin{equation}\label{equ28}
F_{SFG}^{Q}(\omega_l,t_{p})=\frac{8 t_{p}^2n_{sv}  \epsilon}{1+2 n_{t h}-4 n_{sv} \epsilon}\approx\frac{8 t_{p}^2n_{sv}  \epsilon}{1+2 n_{t h}},
\end{equation}
where the approximation holds for $n_{sv} \epsilon \ll 1$. This QFI with the approximation can also be derived from the mean values in Eq. (\ref{equ24}) with an approximated covariance matrix
\begin{equation}\label{equ34}
C^{\prime}_{SFG}\approx\frac{1}{8}\left(\begin{array}{cc}
1+2 n_{t h} & 0 \\
0 & 1+2 n_{t h}
\end{array}\right),
\end{equation}
where the information of the beat in the covariance matrix in Eq. (\ref{equ25}) is omitted. 

The Eqs. (\ref{equ27}) and (\ref{equ28}) suggest that, in the limit $\epsilon \ll 1$ and $n_{s v} \ll 1$ the QFI of the mixed state remains the same before and after the SFG process, i.e., 
$F_{SFG}^{Q}(\omega_l,t_{p})=F_{sv-th}^{Q}(\omega_l,t_{p})$, in the presence of strong thermal noise $n_{th}\gg1$. In contrast, for weak thermal noise $n_{th}\ll 1$, the QFI doubles after the SFG process, yielding $F_{SFG}^{Q}(\omega_l,t_{p})=2F_{sv-th}^{Q}(\omega_l,t_{p})$, corresponding to a 3 dB improvement. Furthermore, when $n_{coh} = n_{sv}$, Eqs. (\ref{equ27}) and (\ref{equ28}) indicate that, in the limit $\epsilon \ll 1$ and $n_{s v} \ll 1$, the QFI of the two-path-mode squeezed vacuum state with the SFG process is twice that of the coherent state, independent of the thermal noise level.

\section{The classical limit of estimating beat information for Gaussian states}

Corresponding to the quantum CRB, the classical CRB establishes a lower bound on the variance of an unknown parameter $x$ within a quantum state $\hat{\rho}(x)$ for a given detection strategy:
\begin{equation}
E[(\hat{x}-x)^{2}]=\delta^{2} \hat{x} \geqslant \frac{1}{\nu } F^{-1} _{C}(x),
\end{equation}
Here $\nu$ represents the number of times the procedure is repeated, $\hat{x}$ is the unbiased estimators of a $x$, $F_{C}(x)$ is the classical Fisher information (CFI) satisfying $F_{C}(x)\leq F_{Q}(x)$. For a specific set of positive operator valued measurements (POVM) $\{ \Pi_{y} \}$ with continuous variable detection outcome $y$, the classical Fisher information $F_{C}(x)$ is given as
\begin{equation}
F_C(x)=-\int^{\infty}_{-\infty}  p(y\mid x) \frac{\partial^2 \ln p(y \mid x)}{\partial^{2} x}dy,
\end{equation}
where $p(y \mid x)=\operatorname{Tr}\left[\hat{\rho}(x)\Pi_{y}\right]$ is a continuous probability distribution.

For Gaussian states with frequency modulation, the beat information can be extracted by measuring the in-phase component $\hat{I}(t_{p})$ and the quadrature component $\hat{Q}(t_{p})$ in the time domain. In this context, the continuous probability distribution of the measurement outcomes can be described by the Wigner function \cite{nielsen2023deterministic} defined in Eq. (\ref{equ29}). However, because $[\hat{I}(t_{p}),\hat{Q}(t_{p})]\neq 0$, the in-phase and quadrature components cannot be simultaneously optimized for measurement unless additional vacuum fluctuation is introduced, as discussed in Sec. \ref{SMQHD}. Consequently, in some cases, QHD may not represent the optimal detection strategy, resulting in $F_{C}(x)\neq F_{Q}(x)$, as shown below.

\subsection{The classical limit of quantum heterodyne detection}

For the frequency-modulated coherent state shown in Eq. (\ref{equ16}) mixed with a multi-frequency-mode CW thermal state shown in Eq. (\ref{equ17}), the mean values and the covariance matrix of in-phase component $\hat{I}_{S}(t_{p})$ and the quadrature component $\hat{Q}_{S}(t_{p})$ of the return state $\rho_{coh-th}$ provided by QHD are given as 
\begin{equation}\label{equ43}
\left\langle I_{R}(t_{p})\right\rangle_{coh-th}=\sqrt{\epsilon}\alpha^{\prime} \cos \left(\omega_l t_{p}+\phi_l\right),\quad \left\langle Q_{R}(t_{p})\right\rangle_{coh-th}=\sqrt{\epsilon}\alpha^{\prime} \sin \left(\omega_l t_{p}+\phi_l\right),
\end{equation}
and
\begin{equation}\label{equ44}
C^{\prime}_{coh-th}=C_{vac}+\epsilon C_{coh}+(1-\epsilon)C_{th},
\end{equation}
where an additional vacuum fluctuation from the image-band mode is introduced into the covariance matrix, compared to Eqs. (\ref{equ30}) and (\ref{equ31}).
The Gaussian distribution, characterized by these mean values and covariance matrix, results in the following CFI
\begin{equation}
F_{coh-th}^{QHD}(\omega_{l},t_{p})=\frac{2 t^2n_{coh} \epsilon}{1+n_{t h}(1-\epsilon)}\approx \frac{2 t_{p}^2n_{coh} \epsilon}{1+n_{t h}}
\end{equation}
where the approximation hold for $\epsilon \ll 1$. This suggests that for a weak reflective target $\epsilon \ll 1$ with strong thermal noise $n_{th} \gg 1$, QHD is the optimal method for the mixed state $\rho_{coh-th}$, yielding $F_{coh-th}^{QHD}(\omega_{l},t_{p}) = F_{coh-th}^{Q}(\omega_{l},t_{p})$. For weak thermal noise $n_{th} \ll1$,  QHD is not longer the optimal method for the mixed state $\rho_{coh-th}$ since that $F_{coh-th}^{QHD}(\omega_{l},t_{p}) = F_{coh-th}^{Q}(\omega_{l},t_{p})/2$.

For the two-path-mode squeezed vacuum state with frequency modulation shown in Eq. (\ref{equ18}) mixed with a multi-frequency-mode CW thermal state shown in Eq. (\ref{equ17}), QHD can be independently applied in both signal and idler modes of the return state $\rho_{sv-th}$. In this case, the mean values and the covariance matrix of in-phase and quadrature components in the time domain of the return state $\rho_{sv-th}$ in continuous time limit are given as
\begin{equation}
\left\langle I_{R}(t_{p})\right\rangle_{sv-th}=0,\quad \left\langle Q_{R}(t_{p})\right\rangle_{sv-th}=0, \quad \left\langle I_{I}(t_{p})\right\rangle_{sv-th}=0,\quad \left\langle Q_{I}(t_{p})\right\rangle_{sv-th}=0,
\end{equation}
and
\begin{equation}
C^{\prime}_{sv-th}=C_{vac}\otimes I_{2}+ C_{sv-th},
\end{equation}
where $I_{2}$ is the identity matrix defined in Eq. (\ref{equ35}). Note that, since the vacuum fluctuations of the image-band for the return mode and idler mode are independent, they are introduced only in the diagonal elements of the covariance matrix. The Gaussian distribution, characterized by these mean values and the covariance matrix, results in the following CFI:
\begin{equation}
F_{sv-th}^{QHD}(\omega_{l},t)=\frac{2 t_{p}^2n_{sv} \epsilon}{1+n_{t h}(1-\epsilon)},
\end{equation}
which is the same as $F_{coh-th}^{QHD}(\omega_{l},t_{p})$ for $n_{coh}=n_{sv}$. Thus, the performance is identical for $\rho_{sv-th}$ and $\rho_{coh-th}$ when QHD is used.

Further, if the signal and idler light of the mixed state $\rho_{sv-th}$ undergo an additional SFG process, in the limit of $\epsilon \ll 1$ and $n_{s v} \ll 1$, the mean values and the covariance matrix of in-phase component $\hat{I}_{P}(t_{p})$ and the quadrature component $\hat{Q}_{P}(t_{p})$ of the up-conversion state $\rho_{SFG}$ provided by QHD are given as
\begin{equation}\label{equ46}
\left\langle I_{P}(t_{p})\right\rangle_{SFG}=-\epsilon_{s}\sqrt{\epsilon n_{s v}} \cos \left(\omega_l t_{p}+\phi_l\right),\quad \left\langle Q_{P}(t_{p})\right\rangle_{SFG}=-\epsilon_{s}\sqrt{\epsilon n_{s v}} \sin \left(\omega_l t_{p}+\phi_l\right),
\end{equation}
and
\begin{equation}\label{equ47}
C^{\prime\prime}_{SFG}=C_{vac}+ \epsilon_s^2 C^{\prime}_{SFG},
\end{equation}
which is similar to Eq. (\ref{equ43}) and Eq. (\ref{equ44}). Then, the Gaussian distribution, characterized by these mean values and the covariance matrix, results in the following CFI
\begin{equation}\label{equ45}
F_{SFG}^{QHD}(\omega_{l},t_{p})=\frac{8t_{p}^2  \epsilon n_{s v} \epsilon_s^2}{2+\left(1+2 n_{t h}\right) \epsilon_s^2},
\end{equation}
where $\epsilon_s^2$ can not be omitted. This suggests that for a weak reflective target 
$\epsilon \ll 1$ with $\epsilon_s=\sqrt{2}$ and $n_{sv}\ll1$, QHD is the optimal method for the mixed state $\rho_{SFG}$, yielding $F_{SFG}^{QHD}(\omega_{l},t_{p}) = F_{SFG}^{Q}(\omega{l},t_{p})$. This conclusion can be further extended to the region of $\epsilon_s\gg\sqrt{2}$. It is worth noting that, $F_{SFG}^{HD}(\omega{l},t_{p})=F_{sv-th}^{Q}(\omega{l},t_{p})$ in the region of $\epsilon^{2}_s n_{th}\gg1$ where $\epsilon_s\ll\sqrt{2}$.

\subsection{The classical limit of quantum heterodyne detection with squeezed vacuum input}

\begin{figure}[h]
\centering
\includegraphics[width=0.5\linewidth]{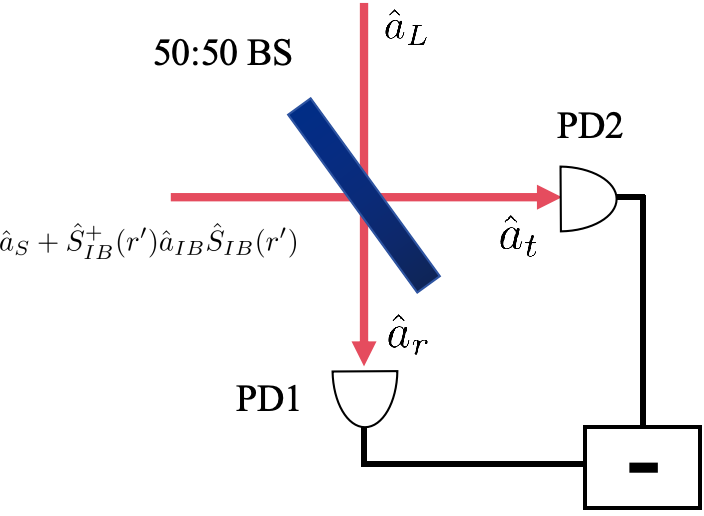}
\caption{The sketch of balanced heterodyne detection with squeezed vacuum state of the image band mode. Here $\hat{S}_{IB}(r^{\prime})=\exp[r^{\prime}(\hat{a}_{IB}^{2}-\hat{a}_{IB}^{\dag2})/2]$ is squeezing operator of image band mode.}
\label{fig:supp4}
\end{figure}

As shown in Eq. (\ref{equ45}), the CFI of QHD for the return state with SFG strongly depends on the strength of the SFG process, $\epsilon_s$, which is typically small. For small $\epsilon_s$ and low background noise ($n_{th} \ll 1$), the signal of the up-conversion mode is modified by $\epsilon_s$, as shown in Eq. (\ref{equ46}), while the noise is dominated by the vacuum fluctuations of the image band, as shown in Eq. (\ref{equ47}), which leads to a poor SNR. One potential solution to this problem is to mitigate vacuum fluctuations by applying squeezing to the image band mode. This can be achieved by introducing a squeezed vacuum state of the image band mode, which is combined with the up-conversion mode for QHD, that is squeezed vacuum injection (SVI), as illustrated in Fig. \ref{fig:supp4}.

For the in-phase component squeezing, the covariance matrix of the squeezed vacuum state for the image band mode is expressed as \cite{weedbrook2012gaussian}:
 \begin{equation}
C_{sv-IB}=\frac{1}{4}\left(\begin{array}{ll}
e^{-2r^{\prime}} &\quad 0 \\
0 &\quad e^{2r^{\prime}}
\end{array}\right),
\end{equation}
where $r^{\prime}$ is the squeezing parameter. Since the noise in the quadrature component increases as the noise in the in-phase component is squeezed, we focus solely on the measurement of the in-phase component of the up-conversion mode.
In this case, the probability distribution obtained by measuring the in-phase component of the up-conversion state $\rho_{SFG}$ is given as:
\begin{equation}
W\left(I_P(t_p)\mid \omega_{l}\right)=\frac{1}{2 \pi \sqrt{  c_{SFG}} } \mathrm{e}^{-\frac{\left(I_P(t_p)-\langle I_P(t_p)\rangle_{SFG}\right)^{2}}{2  c_{SFG}}},
\end{equation}
where
\begin{equation}
\left\langle I_P(t_p)\right\rangle_{SFG}=\epsilon_s\left\langle I^{\prime}_P(t_p)\right\rangle_{SFG}=-\epsilon_s\sqrt{\epsilon n_{s v}} \cos \left(\omega_l t_p +\phi_l\right),\quad c_{SFG}=\epsilon_s^2c^{\prime}_{SFG}=\frac{1}{8}\epsilon_s^2(1+2 n_{t_p h})+\frac{1}{4}e^{-2r^{\prime}},
\end{equation}
which is given in Eq. (\ref{equ24}) and Eq. (\ref{equ34}) with additional squeezed vacuum noise. Then, in the limit of $\epsilon \ll 1$ and $n_{s v} \ll 1$, the CFI of the beat frequency for QHD of this up-conversion state $\rho_{SFG}$ is then given by
\begin{equation}\label{equ48}
\begin{aligned}
F_{SFG}^{HD-SVI}(\omega_{l},t_p)&=-\int^{\infty}_{-\infty}  W\left(I_P(t_p)\mid \omega_{l}\right) \frac{\partial^2 \ln W\left(I_P(t_p)\mid \omega_{l}\right)}{\partial^{2} \omega_{l}}d I_P(t)\\
&=\frac{8t_p^2  \epsilon n_{s v} \epsilon_s^2 e^{2r^{\prime}}}{2+\left(1+2 n_{t h}\right) \epsilon_s^2 e^{2r^{\prime}}}\sin^{2}(\omega_{l} t+\phi_{l})\approx \frac{4t_p^2  \epsilon n_{s v} \epsilon_s^2 e^{2r^{\prime}}}{2+\left(1+2 n_{t h}\right) \epsilon_s^2 e^{2r^{\prime}}},
\end{aligned}
\end{equation}
where the approximation hold for $\omega_{l}\gg1/T_{m}$ when considering the CFI of the entire modulation period,
\begin{equation}
F_{SFG}^{QHD-SVI}(\omega_{l})=\int^{t_{0}+T_m}_{t_{0}} F_{SFG}^{QHD-SVI}(\omega_{l},t) dt,
\end{equation}
while, in the continuous-time limit $t_p\rightarrow t$ for $\Delta\omega_{B}\gg\Delta\omega$. Here, $t_{0}$ represents the initial detection time. Comparing Eq. (\ref{equ48}) with Eq. (\ref{equ45}), the equivalent strength of the SFG process in Eq. (\ref{equ48}) is enhanced by a factor of $e^{2r^{\prime}}$, while its total CFI is reduced by half. As such, for weak background noise $n_{th}\ll1$ with $e^{2r^{\prime}}\epsilon_s^2\gg1$, $F_{SFG}^{QHD-SVI}(\omega_{l},t_p)=2F_{coh-sv}^{QHD}(\omega_{l},t_p)$ with 3dB improvement, while $F_{SFG}^{QHD-SVI}(\omega_{l},t_p)=F_{coh-sv}^{QHD}(\omega_{l},t_p)$ for strong background noise $n_{th}\gg1$ with $e^{2r^{\prime}}\epsilon_s^2\gg1$.

\bibliography{bibtex}